\documentclass[journal]{IEEEtran}

\usepackage{graphicx}
\usepackage{subfigure}
\usepackage{subeqnarray}
\usepackage{bm}
\usepackage{amsfonts}
\usepackage{amsthm}
\usepackage{amssymb}
\usepackage{makecell}
\usepackage{calc}
\usepackage{color}
\usepackage{mathrsfs}
\usepackage{flushend}
\usepackage{url}

\usepackage[outdir=./]{epstopdf}

\usepackage{cite}

\usepackage{booktabs}

\ifCLASSINFOpdf
\else
\fi
\usepackage{amsmath}
\usepackage{amsthm}

\usepackage{algorithm}
\usepackage{algorithmic}
\ifCLASSOPTIONcompsoc
 \usepackage[caption=false,font=normalsize,labelfont=sf,textfont=sf]{subfig}
\else
 \usepackage[caption=false,font=footnotesize]{subfig}
\fi
\newcommand{\tabincell}[2]{\begin{tabular}{@{}#1@{}}#2\end{tabular}}

\begin{document}
\captionsetup[figure]{labelformat={default},labelsep=period,name={Fig.}}
\title{Secure Radar-Communication Systems with Malicious Targets: Integrating Radar, Communications and Jamming Functionalities}
%
%
%

\author{Nanchi Su,~\IEEEmembership{Student~Member,~IEEE,}
        Fan Liu,~\IEEEmembership{Member,~IEEE,}
        and Christos Masouros,~\IEEEmembership{Senior~Member,~IEEE}
\thanks{This paper was supported in part by the European Union's Horizon 2020 research and innovation programme under the Marie Sk\l odowska-Curie Grant Agreement No. 793345, in part by the Engineering and Physical Sciences Research Council (EPSRC) of the UK Grant number EP/R007934/1 and EP/S026622/1, in part by the UK MOD University Defence Research Collaboration
(UDRC) in Signal Processing, and in part by the China Scholarship Council (CSC).}
\thanks{N. Su, F. Liu and C. Masouros are with the Department of Electronic and Electrical Engineering, University College London, London WC1E 7JE, U.K. (e-mail:
nanchi.su.18@ucl.ac.uk; fan.liu@ucl.ac.uk; chris.masouros@ieee.org).}
}

\maketitle

\begin{abstract}
    This paper studies the physical layer security in a multiple-input-multiple-output (MIMO) dual-functional radar-communication (DFRC) system, which communicates with downlink cellular users and tracks radar targets simultaneously. Here, the radar targets are considered as potential eavesdroppers which might eavesdrop the information from the communication transmitter to legitimate users. To ensure the transmission secrecy, we employ artificial noise (AN) at the transmitter and formulate optimization problems by minimizing the signal-to-interference-plus-noise ratio (SINR) received at radar targets, while guaranteeing the SINR requirement at legitimate users. We first consider the ideal case where both the target angle and the channel state information (CSI) are precisely known. The scenario is further extended to more general cases with target location uncertainty and CSI errors, where we propose robust optimization approaches to guarantee the worst-case performances. Accordingly, the computational complexity is analyzed for each proposed method. Our numerical results show the feasibility of the algorithms with the existence of instantaneous and statistical CSI error. In addition, the secrecy rate of secure DFRC system grows with the increasing angular interval of location uncertainty.
\end{abstract}


%
\IEEEpeerreviewmaketitle

\section{Introduction}
%
%
%
%
\IEEEPARstart{T}{he} increasing spectrum congestion has intensified the efforts in dynamic spectrum licensing and soon spectrum is to be shared between radar and communication applications. Govermental organizations such as the US Department of Defence (DoD) have a documented requirement of releasing 865 MHz to support telemetry by the year of 2025, but only 445 MHz is available at present \cite{oyediran2015spectrum}. As a result, the operating frequency bands of communication and radar are overlapped with each other \cite{li2016optimum}, which leads to mutual interference between two systems. Furthermore, both systems have been recently given a common spectrum portion by the Federal Communication Commission (FCC) \cite{1542627,kim2015design,staple2004end}. To enable the efficient usage of the spectrum, research efforts are well underway to address the issue of communication and radar spectrum sharing (CRSS).
\\\indent Aiming for realizing the spectral coexistence of individual radar and communication systems, several interference mitigation techniques have been proposed in \cite{sodagari2012projection, turlapaty2014joint, li2017joint, 8288677, liu2017robust, liu2018mimo}. As a step further, dual-functional radar-communication (DFRC) system that is capable of realizing not only the spectral coexistence, but also the shared use of the hardware platform, has been regarded as a promising research direction \cite{8386661, liu2018dual, zhou2018optimal, hassanien2016dual}. It is noteworthy that the DFRC technique has already been widely explored in numerous civilian and military applications, including 5G vehicular network \cite{va2016millimeter}, WiFi based indoor positioning \cite{lim2007real}, low-probability-of-intercept (LPI) communication \cite{dillard2003cyclic} as well as the advanced multi-function radio frequency concept (AMRFC) \cite{mccormick2017simultaneous}.
\\\indent In the DFRC system, radar and communication functionalities are realized by a well-designed probing waveform that carries communication signalling and data. Evidently, this operation implicates security concerns, which are largely overlooked in the relevant DFRC literature. It is known that typical radar requires to focus the transmit power towards the directions of interest to obtain a good estimation of the targets. Nevertheless, in the case of DFRC transmission, critical information embedded in the probing waveform could be leaked to the radar targets, which might be potential eavesdroppers at the adversary's side. To this end, it is essential to take information security into consideration for the DFRC design. In the communication literature, physical layer security has been widely investigated, where the eavesdroppers' reception can be crippled by exploiting transmit degrees of freedom (DoFs) \cite{liao2011qos}. MIMO secrecy capacity problems were considered in \cite{shafiee2007towards,oggier2007secrecy,ekrem2011secrecy}. Besides, another meaningful technique for enabling physical layer secrecy was presented in \cite{liao2011qos, goel2008guaranteeing}, namely artificial noise (AN) aided transmission. Furthermore, the AN generation algorithms studied in \cite{negi2005secret,zhang2013design} were with the premise of publicly known channel state information (CSI) in a fading environment. Moreover, some concurrent AN-aided studies employed cooperative jammers to improve secure communication \cite{5708173,chu2015secrecy}.
\\\indent Given the dual-functional nature of the DFRC systems, the secrecy issue can be addressed on the aspect of either radar or communication. From the perspective of the radar system, existing works focus on the radar privacy maintenance \cite{li2017joint, vaka2016location, dimas2017spectrum}. A functional architecture was presented in \cite{li2017joint} for the control center aiming at coordinating the cooperation between radar and communication while maintaining the privacy of the radar system. In \cite{vaka2016location}, obfuscation techniques have been proposed to counter the inference attacks in the scenario of spectrum sharing between military radars and commercial communication systems. Besides, the work of \cite{dimas2017spectrum} showed the probability for an adversary to infer radar's location by exploiting the communication precoding matrices. On the other hand, the works of \cite{deligiannis2018secrecy,chalise2018performance} have studied the secrecy problems from the viewpoint of communications. In \cite{deligiannis2018secrecy}, the MIMO radar transmits two different signals simultaneously, one of which is embedded with desired information for the legitimate receiver, the other one consists of false information to confuse the eavesdroppers. Both of the signals are used to detect the target. Several optimization problems were presented, including secrecy rate maximization, target return signal-to-interference-plus-noise ratio (SINR) maximization and transmit power minimization. Then, a unified joint system of passive radar and communication systems was considered in \cite{chalise2018performance}, where the communication receivers might be eavesdropped by the target of passive radar. To guarantee the secrecy of legitimate user in the communication system, the optimization problem was designed to maximize the SINR at the passive radar receiver (RR) while keeping the secrecy rate above a certain threshold. While the aforementioned approaches are well-designed by sophisticated techniques, the AN-aided physical layer security remains to be explored for the DFRC systems under practical constraints.
\\\indent To the best of our knowledge, most of the present works regarding secure transmission in DFRC system rely on the assumption of precisely known channel state information (CSI) at the transmitter. To address the beamforming design in a general context, we take the imperfect CSI into account in our work, which includes instantaneous and statistical CSI with norm-bounded errors. Moreover, the well-known S-procedure and Lagrange dual function have been adopted to reformulate the optimization problem, which can be solved by Semidifinite Relaxation (SDR) approach. In addition to the CSI issues, we also explore the radar-specific target uncertainty, where we employ a robust adaptation technique for target tracking.
\\\indent Accordingly, in this paper, we propose several optimization problems aiming at ensuring information transmission security of the DFRC system. To be specific, we consider a MIMO DFRC base station (BS) that is serving multiple legitimate users while detecting targets. It should be noted that these targets are assumed to be potential eavesdroppers. Moreover, spatially focused AN is employed in our methods. Throughout the paper, we aim to minimize the SINR at the target while ensuring the SINR at each legitimate user. Within this scope, we summarize our contributions as follows:
\begin{itemize}
    \item We first consider the ideal scenario under the assumptions of perfect CSI and known precise location of targets. The beampattern is formed by approaching to a given benchmark radar beampattern. By doing so, the formulated optimization problem can be firstly recast as Fractional programming (FP) problem \cite{8314727}, and then solved by the SDR.
    \item We investigate the problem under the practical condition of target location uncertainty, where we formulate a beampattern with a given angular interval that the targets might fall into.
    \item We impose the imperfect communication CSI to the optimization in addition to the above constraints, where worst-case FP problems are formulated to minimize the maximum SINR at the target with bounded CSI errors.
    \item We consider the statistical CSI, which is more practical due to significantly reduced feedback requirements \cite{wajid2009robust}. To tackle this scenario, we further formulate the eavesdropper SINR minimization problem considering the error bound of statistical CSI.
    \item We derive the computational complexity for each proposed algorithm.
\end{itemize}

This paper is organized as follows. Section \uppercase\expandafter{\romannumeral2} gives the system model. The optimization problems based on perfect CSI are addressed in Section \uppercase\expandafter{\romannumeral3} and \uppercase\expandafter{\romannumeral4} for precise location and uncertain direction of targets, respectively. In Section \uppercase\expandafter{\romannumeral5} and \uppercase\expandafter{\romannumeral6}, more general context of imperfect CSI is considered, which addresses issues with imperfect CSI under norm-bounded and statistical errors, respectively. Section \uppercase\expandafter{\romannumeral7} provides numerical results, and Section \uppercase\expandafter{\romannumeral8} concludes the paper.
\\\indent\emph{Notations}: Unless otherwise specified, matrices are denoted by bold uppercase letters (i.e., $\mathbf{H}$), vectors are represented by bold lowercase letters (i.e., $\mathbf{x}$), and scalars are denoted by normal font (i.e., $\alpha$). Subscripts indicate the location of the entry in the matrices or vectors (i.e., $s_{i,j}$ and $l_n$ are the $(i,j)$-th and the \emph{n}-th element in $\mathbf{S}$ and $\mathbf{l}$, respectively). $\operatorname{tr}\left(\cdot\right)$ and $\operatorname{vec}\left(\cdot\right)$ denote the trace and the vectorization operations. $\left(\cdot\right)^T$, $\left(\cdot\right)^H$, $\left(\cdot\right)^*$ and $\left(\cdot\right)^\dag$ stand for transpose, Hermitian transpose, complex conjugate and Moore-Penrose pseudo-inverse of the matrices, respectively. $\operatorname{diag}\left(\cdot\right)$ represents the vector formed by the diagonal elements of the matrices and ${\text{rank}}\left(  \cdot  \right)$ is rank operator. $\left\| \cdot\right\|$, $\left\| \cdot\right\|_{\infty}$ and $\left\| \cdot\right\|_F$ denote the $l_2$ norm, $l_{\infty}$ and the Frobenius norm respectively. $\mathbb{E}\left\{ \cdot  \right\}$ denotes the statistical expectation. ${\left[  \cdot  \right]^{\text{ + }}}$ denotes $\max \left\{ { \cdot ,0} \right\}$.

\begin{figure}
    \centering
    \includegraphics[width=1.0\columnwidth]{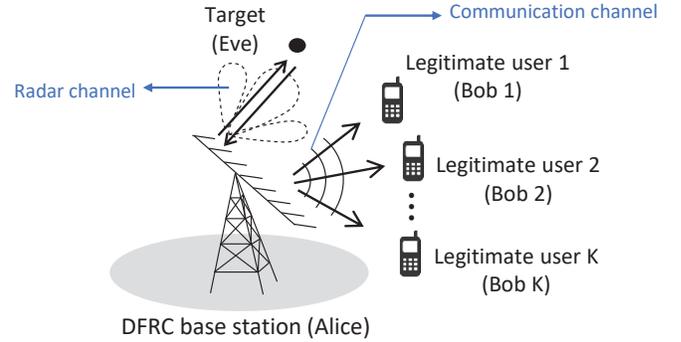}
    \captionsetup{font={footnotesize}}
    \caption{Dual-functional Radar-Communication system detecting target which comprise a potential eavesdropper.}
    \label{fig.1}
\end{figure}

\section{System Model}
We consider a dual-functional MIMO DFRC system, which consists of a DFRC base station, legitimate users and target which is a potential eavesdropper, as shown in Fig. 1. The DFRC system is equipped with uniform linear array (ULA) of \emph{N} antennas, serving \emph{K} single-antenna users, while detecting a single point-like target. For convenience, the multi-antenna transmitter, the legitimate users and the target will be referred as Alice, Bobs and Eve respectively.
\subsection{Signal Model}
In the scenario shown in Fig. 1, the DFRC base station Alice intends to send confidential information to single-antenna legitimate users, i.e. Bobs, with the presence of the potential eavesdropper, i.e. Eve. The received symbol vector at Bobs can be modeled as
\begin{equation}\label{eq1}
    {\mathbf{y} = {\mathbf{H}}\mathbf{x} + \mathbf{z}}\,
\end{equation}
where ${\mathbf{H}} = {\left[ {{{\mathbf{h}}_1},{{\mathbf{h}}_2}, \cdots ,{{\mathbf{h}}_K}} \right]^T}\in {\mathbb{C}^{K\times N}}$ is the channel matrix, ${\mathbf{x}}\in {\mathbb{C}^{N}}$ is the transmitted signal vector,  ${\mathbf{z}}$ is the noise vector, with ${{\mathbf{z}}}\sim\mathcal{C}\mathcal{N}\left( {0,{\sigma^2}{\mathbf{I}}_N} \right)$.
\\\indent Consider AN-aided transmit beamforming, the transmit vector ${\mathbf{x}}$ can be written as
\begin{equation}\label{eq2}
    {\mathbf{x} = \mathbf{W} \mathbf{s} + \mathbf{n}}\,
\end{equation}
where ${\mathbf{s}}\in {\mathbb{C}^{K}}$ is the desired symbol vector of Bobs, where we assume $\mathbb{E}\left[ {{\mathbf{s}}}\mathbf{s}^{H} \right]={\mathbf{I}}$, ${\mathbf{W}} = \left[ {{{\mathbf{w}}_1},{{\mathbf{w}}_2}, \cdots ,{{\mathbf{w}}_K}} \right] \in {\mathbb{C}^{N\times K}}$ is the beamforming matrix, ${\mathbf{n}}$ is an artificial noise vector generated by Alice to avoid leaking information to Eves. It is assumed that ${{\mathbf{n}}}\sim\mathcal{C}\mathcal{N}\left( {0,{{\mathbf{R}}_N}} \right)$. Additionally, we assume that the desired symbol vector $\mathbf{s}$ and the artificial noise vector $\mathbf{n}$ are independent with each other.
\\\indent According to \cite{8288677}, it is presumed that the above signal is used for both radar and communication operations, where each communication symbol is considered as a snapshot of a radar pulse. Then, the covariance matrix of radar system can be given as
\begin{equation}\label{eq3}
    {{\mathbf{R}}_X} = \mathbb{E}\left[{\mathbf{x}}{{\mathbf{x}}^H}\right] = \sum\limits_{i = 1}^K {{{\mathbf{W}}_i} + {{\mathbf{R}}_N}},
\end{equation}
where ${{\mathbf{W}}_{i}}\triangleq{{\mathbf{w}}_{i}}\mathbf{w}_{i}^{H}$. Then, the beampattern can be expressed as
\begin{equation}\label{eq4}
    {{P}_{bp}}={{\mathbf{a}}^{H}}\left( \theta  \right){{\mathbf{R}}_{X}}\mathbf{a}\left( \theta  \right),
\end{equation}
where $\theta $ is the angle of target, $\mathbf{a}\left( \theta  \right)={{\left[ \begin{matrix}
   1 & {{e}^{j2\pi \Delta \sin \left( \theta  \right)}} & \cdots  & {{e}^{j2\pi \left( N-1 \right)\Delta \sin \left( \theta  \right)}}  \\
\end{matrix} \right]}^{T}}\in {{\mathbb{C}}^{N\times 1}}$ denotes the steering vector of the transmit antenna array, and $\Delta$ is the interval between adjacent antennas being normalized by the wavelength.
\subsection{Metrics}
To evaluate the performance of the system, we define a number of performance metrics in this subsection. Initially, based on the aforementioned system model, the SINR of the \emph{i}-th user can be written as
\begin{equation}\label{eq5}
\begin{aligned}
    {{\text{SINR}}_{i}}&=\frac{\mathbb{E}\left[ {{\left| \mathbf{h}_{i}^{T}{{\mathbf{w}}_{i}}{{\mathbf{s}}} \right|}^{2}} \right]}{\sum\nolimits_{k\ne i, k=1}^{K}{\mathbb{E}\left[ {{\left| \mathbf{h}_{i}^{T}{{\mathbf{w}}_{k}}{{\mathbf{s}}} \right|}^{2}} \right]+\mathbb{E}\left[{{\left| \mathbf{h}_{i}^{T}{{\mathbf{n}}} \right|}^{2}}\right]+\sigma^{2}}}\\&=\frac{{{\mathbf{h}}_i^T{{\mathbf{W}}_i}{\mathbf{h}}_i^*}}{{\sum\nolimits_{k \ne i,k = 1}^K {\left( {{\mathbf{h}}_i^T{{\mathbf{W}}_k}{\mathbf{h}}_i^*} \right)}  + \left( {{\mathbf{h}}_i^T{{\mathbf{R}}_N}{\mathbf{h}}_i^*} \right) + {\sigma ^2}}},
\end{aligned}
\end{equation}
where ${{\mathbf{n}}_{i}}$ is the AN of \emph{i}-th user.
\\\indent Equation (5) can be simplified
\begin{equation}\label{eq6}
{{\text{SINR}}_{i}}=\frac{\text{tr}\left( \mathbf{h}_{i}^{*} \mathbf{h}_{i}^{T}{{\mathbf{W}}_{i}} \right)}{\sum\nolimits_{k\ne i, k=1}^{K}{\text{tr}\left( \mathbf{h}_{i}^{*}\mathbf{h}_{i}^{T}{{\mathbf{W}}_{k}} \right)}+\text{tr}\left( \mathbf{h}_{i}^{*}\mathbf{h}_{i}^{T}{{\mathbf{R}}_{N}} \right)+\sigma^{2}}.
\end{equation}
The achievable transmission rate of legitimate users is given as
\begin{equation}\label{eq7}
    {R_{{C_i}}} = {\log _2}\left( {1 + {\text{SIN}}{{\text{R}}_i}} \right).
\end{equation}
\\\indent Likewise, based on the given signal model in (\ref{eq3}) and (\ref{eq4}), SINR at Eve can be given as \cite{doi:10.1002/ett.4460100604}
\begin{equation}\label{eq8}
    \text{SIN}{{\text{R}}_{E}}=\frac{{{\left| \alpha  \right|}^{2}}{{\mathbf{a}}^{H}}\left( \theta  \right)\sum\nolimits_{i=1}^{K}{{{\mathbf{W}}_{i}}}\mathbf{a}\left( \theta  \right)}{{{\left| \alpha  \right|}^{2}}{{\mathbf{a}}^{H}}\left( \theta  \right){{\mathbf{R}}_{N}}\mathbf{a}\left( \theta  \right)+\sigma^{2}},
\end{equation}
where $\alpha$ represents the propagation loss in radar system. The achievable transmission rate of Eve can be expressed as
\begin{equation}\label{eq9}
    {R_E} = {\log _2}\left( {1 + {\text{SIN}}{{\text{R}}_E}} \right).
\end{equation}
 Additionally, the transmit power is expressed as
\begin{equation}\label{eq10}
    {{P}_{t}}=\text{tr}({{\mathbf{R}}_{X}}).
\end{equation}
Given the achievable transmission rates of Bobs and Eve, the achievable secrecy rate of the system is defined as \cite{6626661}
\begin{equation}\label{eq11}
   {\text{SR}} = \mathop {\min }\limits_i \frac{1}{2}{\left[ {{R_{{C_i}}} - {R_E}} \right]^ + }.
\end{equation}
\section{Minimizing SINR of Eve With Premise of Perfect CSI and Target Direction}
In this section, we aim to enhance the secrecy rate by minimizing the SINR of Eve and setting a lower threshold of SINR for the legitimate users, i.e. Bobs. The optimization problem is based on the assumption that the channel information from Alice to Bobs in the communication system is known perfectly. Meanwhile, the precise direction of the detected target is known to the transmitter. We shall relax the above assumptions in the following sections.
\subsection{Problem Formulation}
Let us firstly consider the $\text{SINR}_E$ minimization problem, which should guarantee: a) individual SINR requirement at each legitimate user, b) transmit power budget and c) a desired radar spatial beampattern. Note that an ideal radar beampattern should be obtained before designing the beamforming and artificial noise, which can be generated by solving the following constrained least-squares (LS) problem \cite{8288677,4516997} as an example
\begin{equation}\label{eq12}
\begin{gathered}
  \mathop {\min }\limits_{\eta ,{{\mathbf{R}}_d}}\; {\sum\limits_{m = 1}^M {\left| {\eta {P_d}\left( {{\theta _m}} \right) - {{\mathbf{a}}^H}\left( {{\theta _m}} \right){{\mathbf{R}}_d}{\mathbf{a}}\left( {{\theta _m}} \right)} \right|} ^2} \hfill \\
  s.t.\;\;\;{\text{tr}}\left( {{{\mathbf{R}}_d}} \right) = {P_0}, \hfill \\
  \;\;\;\;\;\;\;\;{{\mathbf{R}}_d} \succeq 0,{{\mathbf{R}}_d} = {\mathbf{R}}_d^H, \hfill \\
  \;\;\;\;\;\;\;\;\eta  \ge 0, \hfill \\
\end{gathered}
\end{equation}
where $\eta$ is a scaling factor, ${P_0}$ represents the transmission power budget, $\left\{ {{\theta _m}} \right\}_{m = 1}^M$ denotes an angular grid covering the detection angular range in $\left[ {{ - }\pi /2,\pi /2} \right]$, ${{\mathbf{a}}\left( {{\theta _m}} \right)}$ denotes steering vector, ${{P_d}\left( {{\theta _m}} \right)}$ is the desired ideal beampattern gain at ${\theta _m}$, ${{{\mathbf{R}}_d}}$ represents the desired waveform covariance matrix.
\\\indent Given a covariance matrix ${{{\mathbf{R}}_d}}$ that corresponds to a well-designed MIMO radar beampattern, the fractional programming optimization problem of minimizing ${{\text{SIN}}{{\text{R}}_E}}$ can be formulated as
 \begin{subequations}\label{eq13}
\begin{align}
  &\mathop {\min }\limits_{{{\mathbf{W}}_i},{{\mathbf{R}}_N}}     \frac{{{\left| \alpha  \right|}^{2}}{{\mathbf{a}}^{H}}\left( \theta_0  \right)\sum\nolimits_{i=1}^{K}{{{\mathbf{W}}_{i}}}\mathbf{a}\left( \theta_0  \right)}{{{\left| \alpha  \right|}^{2}}{{\mathbf{a}}^{H}}\left( \theta_0  \right){{\mathbf{R}}_{N}}\mathbf{a}\left( \theta_0  \right)+\sigma^{2}}, \hfill \\
  &s.t.\;\;\;{\left\| {{{\mathbf{R}}_X} - {{\mathbf{R}}_d}} \right\|^2} \le {\gamma _{bp}}, \hfill \\
  &\;\;\;\;\;\;\;\;{\text{SIN}}{{\text{R}}_i} \ge {\gamma _b}, \forall i, \hfill \\
  &\;\;\;\;\;\;\;\;\text{tr}({{\mathbf{R}}_{X}}) = {P_0}, \hfill \\
  &\;\;\;\;\;\;\;\;{{\mathbf{W}}_i} = {\mathbf{W}}_i^H, {{\mathbf{W}}_i} \succeq 0, \forall i, \hfill \\
  &\;\;\;\;\;\;\;\;{\text{rank}}\left( {{{\mathbf{W}}_i}} \right) = 1, \forall i, \hfill \\
  &\;\;\;\;\;\;\;\;{{\mathbf{R}}_N} = {\mathbf{R}}_N^H,{{\mathbf{R}}_N} \succeq 0,
\end{align}
\end{subequations}
where the constraints ${{\mathbf{W}}_i} = {\mathbf{W}}_i^H, {{\mathbf{W}}_i} \succeq 0, {\text{rank}}\left( {{{\mathbf{W}}_i}} \right) = 1, \forall i,$ are equivalent to constraining ${{\mathbf{W}}_{i}}={{\mathbf{w}}_{i}}\mathbf{w}_{i}^{H}$ \cite{liao2011qos}. $\theta_0$ represents the direction of Eve known at Alice\footnote{The MIMO radar is assumed to be with two working modes including searching and tracking. In the search mode, the radar transmits a spatially orthogonal waveform, which formulates the omni-directional beampattern. Potential targets can be searched via the beampattern. Then, the radar is able to track potential targets via transmitting directional waveforms. Thus, the precise location is available to be known at Alice.}, ${\gamma _{bp}}$ is the pre-defined threshold that constraints the mismatch between designed covariance matrix ${{{\mathbf{R}}_X}}$ and the desired ${{{\mathbf{R}}_d}}$, and finally ${\gamma _{b}}$ denotes the predefined SINR threshold of each legitimate user.
\\\indent First, let us employ the SDR approach by relaxing the optimization problem by omitting the ${\text{rank}}\left( {{{\mathbf{W}}_i}} \right) = 1$ constraint in (13f), which can be written as
\begin{subequations}\label{eq14}
\begin{align}
  &\mathop {\min }\limits_{{{\mathbf{W}}_i},{{\mathbf{R}}_N}}     \frac{{{\left| \alpha  \right|}^{2}}{{\mathbf{a}}^{H}}\left( \theta_0  \right)\sum\nolimits_{i=1}^{K}{{{\mathbf{W}}_{i}}}\mathbf{a}\left( \theta_0  \right)}{{{\left| \alpha  \right|}^{2}}{{\mathbf{a}}^{H}}\left( \theta_0  \right){{\mathbf{R}}_{N}}\mathbf{a}\left( \theta_0  \right)+\sigma^{2}}, \hfill \\
  &s.t.\;\;\;{\left\| {{{\mathbf{R}}_X} - {{\mathbf{R}}_d}} \right\|^2} \le {\gamma _{bp}}, \hfill \\
  &\;\;\;\;\;\;\;\;{\text{SIN}}{{\text{R}}_i} \ge {\gamma _b}, \forall i, \hfill \\
  &\;\;\;\;\;\;\;\;\text{tr}({{\mathbf{R}}_{X}}) = {P_0}, \hfill \\
  &\;\;\;\;\;\;\;\;{{\mathbf{W}}_i} = {\mathbf{W}}_i^H, {{\mathbf{W}}_i} \succeq 0, \forall i, \hfill \\
  &\;\;\;\;\;\;\;\;{{\mathbf{R}}_N} = {\mathbf{R}}_N^H,{{\mathbf{R}}_N} \succeq 0.
\end{align}
\end{subequations}
By noting the fact that problem (14) is still non-convex due to the fractional objective function, we propose in the following an iterative approach to solve the problem efficiently.
\subsection{Efficient Solver}
Following \cite{8314727}, (\ref{eq14}) is single-ratio FP problem, which can be solved by employing the Dinkelbach's transform demonstrated in \cite{dinkelbach1967nonlinear}, where the globally optimal solution can be obtained by solving a sequence of SDPs. To develop the algorithm, we firstly introduce a scaling factor $c = {\text{SIN}}{{\text{R}}_E}$, which is an auxiliary variable. We then define two scaling variables $\operatorname{U}$ and $\operatorname{V}$, which are nonnegative and positive respectively, where $\operatorname{U}={\left| \alpha  \right|^2}{{\mathbf{a}}^H}\left( \theta  \right)\sum\nolimits_{i = 1}^K {{{\mathbf{W}}_i}} {\mathbf{a}}\left( \theta  \right),\forall i$, $\operatorname{V}  = {\left| \alpha  \right|^2}{{\mathbf{a}}^H}\left( \theta  \right){{\mathbf{R}}_N}{\mathbf{a}}\left( \theta  \right) + {\sigma ^2}$. As a result, the FP problem (14) can be reformulated as
\begin{subequations}\label{eq15}
\begin{align}
  &\mathop {\min }\limits_{{{\mathbf{W}}_i},{{\mathbf{R}}_N}} \operatorname{U}  - c\operatorname{V} , \hfill \\
  &s.t.\;\;\;{\left\| {{{\mathbf{R}}_X} - {{\mathbf{R}}_d}} \right\|^2} \le {\gamma _{bp}}, \hfill \\
  &\;\;\;\;\;\;\;\;{\text{SIN}}{{\text{R}}_i} \ge {\gamma _b}, \forall i, \hfill \\
  &\;\;\;\;\;\;\;\;\text{tr}({{\mathbf{R}}_{X}}) = {P_0}, \hfill \\
  &\;\;\;\;\;\;\;\;{{\mathbf{W}}_i} = {\mathbf{W}}_i^H, {{\mathbf{W}}_i} \succeq 0, \forall i, \hfill \\
  &\;\;\;\;\;\;\;\;{{\mathbf{R}}_N} = {\mathbf{R}}_N^H,{{\mathbf{R}}_N} \succeq 0,
\end{align}
\end{subequations}
where $c$ can be iteratively updated by
\begin{equation}\label{eq16}
    c\left[ {t + 1} \right] = \frac{{\operatorname{U}\left[ t \right]}}{{\operatorname{V}\left[ t \right]}},
\end{equation}
where $t$ is the index of iteration. For clarity, we summarize the above in Algorithm 1. According to \cite{8314727}, it is easy to prove the convergence of the algorithm given the non-increasing property of $c$ during each iteration. It is noted that the SDR approach generates an approximated solution to the optimization problem (13) by neglecting the rank-one constraint. Accordingly, eigenvalue decomposition or Gaussian randomization techniques are commonly employed to obtain a suboptimal solution.
\renewcommand{\algorithmicrequire}{\textbf{Input:}}
\renewcommand{\algorithmicensure}{\textbf{Output:}}
\begin{algorithm}
\caption{Alogrithm for solving FP problem (\ref{eq14})}
\label{alg:1}
\begin{algorithmic}
    \REQUIRE ${\mathbf{H}},{\mathbf{a}}\left( \theta_0  \right),{\sigma ^2},\alpha ,{\gamma _{bp}},{\gamma _b},{P_0},ite{r_{max}} \ge 2$
    \ENSURE ${\mathbf{W}}_i^{\left( {iter} \right)},{\mathbf{R}}_N^{\left( {iter} \right)},i = 1, \cdots ,K$
    \STATE 1. Compute ${{\mathbf{R}}_d}$. Reformulate problem (13a) by (14). Set the iteration threshold $\varepsilon > 0$. Initialize ${c^{\left( 0 \right)}},{c^{\left( 1 \right)}},\left| {{c^{\left( 1 \right)}} - {c^{\left( 0 \right)}}} \right| > \varepsilon $.
    \WHILE {$iter \le ite{r_{max}}$ and $\left| {{c^{iter + 1}} - {c^{iter}}} \right| \ge \varepsilon$ }
    \STATE 2. Solve the SDP problem (15).
    \STATE 3. Update $c$ by (16).
    \STATE 4. $ iter = iter + 1$.
    \ENDWHILE
\STATE 6. Obtain approximated solutions by eigenvalue decomposition or Gaussian randomization.
\end{algorithmic}
\end{algorithm}
\subsection{Complexity Analysis}
 In this subsection, the computational complexity of Algorithm 1 is analyzed as follows. Note that SDP problems are commonly solved by the interior point method (IPM) \cite{wang2014outage}, which obtains an $\epsilon$-optimal solution after a sequence of iterations with the given $\epsilon$. In problem (15), it is noted that the constraints are linear matrix inequality (LMI) except for (15b), which is a second-order cone (SOC) \cite{lobo1998applications} constraint. Besides, we note that the solution is required to satisfy the rank-one constraint, the complexity of eigenvalue decomposition\footnote{Eigenvalue decomposition is adopted to obtain a sub-optimal result because of the high complexity of Gaussian randomization.} is then taken into consideration, which is operated at the cost of $\mathcal{O}\left( {\left( {K + 1} \right){N^3}} \right)$ complex multiplications. Thus, we demonstrate the complexity in Table I, where ${N_{iter}}$ represents iteration times. For simplicity, the computational complexity can be given as $\mathcal{O}\left( {\sqrt 2 {N_{iter}}\ln \left( {{1 \mathord{\left/
 {\vphantom {1 \epsilon }} \right.
 \kern-\nulldelimiterspace} \epsilon }} \right){K^{3.5}}{N^{6.5}}} \right)+\mathcal{O}\left( {\left( {K + 1} \right){N^3}} \right)$by reserving the highest order term.
\section{Eve's SINR Minimization With Uncertainty in the Target Direction and Perfect CSI}
In practice, the precise location of the target is difficult to be known at transmitter. In this section, we consider the scenario where a rough estimate of the target angle, instead of its precise counterpart, is available at Alice. Therefore, the following beampattern design aims at achieving both a desired main-beam width covering the possible angle uncertainty interval of the target as well as a minimized sidelobe power in a prescribed region.
\subsection{Problem Formulation}
In this subsection, we consider the case that the angle uncertainty interval of the target is roughly known within the angular interval $\left[ {{\theta _0} - \Delta \theta ,{\theta _0} + \Delta \theta } \right]$. To this end, the target from every possible direction should be taken in to consideration when formulating the optimization problem. Accordingly, the objective is given as the sum of Eve's SINR at all the possible locations as follows. Due to the uncertainty of target location, wider beampattern needs to be formulated towards the uncertain angular interval to avoid missing the target. Inspired by the 3dB main-beam width beampattern design for MIMO radar \cite{4350230}, we propose a scheme aiming at keeping a constant power in the uncertain angular interval, which can be formulated as the following optimization problem
\begin{subequations}\label{eq17}
\begin{align}
  &\mathop {\min }\limits_{{{\mathbf{W}}_i},{{\mathbf{R}}_X}} \sum\limits_{{\theta _m} \in \Phi } {\frac{{{{\left| \alpha  \right|}^2}{{\mathbf{a}}^H}\left( {{\theta _m}} \right)\sum\nolimits_{i = 1}^K {{{\mathbf{W}}_i}} {\mathbf{a}}\left( {{\theta _m}} \right)}}{{{{\left| \alpha  \right|}^2}{{\mathbf{a}}^H}\left( {{\theta _m}} \right){{\mathbf{R}}_N}{\mathbf{a}}\left( {{\theta _m}} \right) + {\sigma ^2}}}}  \hfill \\
  &s.t.\;\;{{\mathbf{a}}^H}\left( {{\theta _0}} \right){{\mathbf{R}}_X}{\mathbf{a}}\left( {{\theta _0}} \right) - {{\mathbf{a}}^H}\left( {{\theta _m}} \right){{\mathbf{R}}_X}{\mathbf{a}}\left( {{\theta _m}} \right) \ge {\gamma _s}, \nonumber \\
  &\;\;\;\;\;\;\;\;\;\;\;\;\;\;\;\;\;\;\;\;\;\;\;\;\;\;\;\;\;\;\;\;\;\;\;\;\;\;\;\;\;\;\;\;\;\;\;\;\;\;\;\;\;\;\;\;\;\;\;\;\;\;\;\;\;\forall {\theta _m} \in \Omega  \hfill \\
  &\;\;\;\;\;\;\;{{\mathbf{a}}^H}\left( {{\theta _k}} \right){{\mathbf{R}}_X}{\mathbf{a}}\left( {{\theta _k}} \right) \le \left( {1 + \alpha } \right){{\mathbf{a}}^H}\left( {{\theta _0}} \right){{\mathbf{R}}_X}{\mathbf{a}}\left( {{\theta _0}} \right),\nonumber \\
  &\;\;\;\;\;\;\;\;\;\;\;\;\;\;\;\;\;\;\;\;\;\;\;\;\;\;\;\;\;\;\;\;\;\;\;\;\;\;\;\;\;\;\;\;\;\;\;\;\;\;\;\;\;\;\;\;\;\;\;\;\;\;\;\;\;\forall {\theta _k} \in \Phi  \hfill \\
  &\;\;\;\;\;\;\;\left( {1 - \alpha } \right){{\mathbf{a}}^H}\left( {{\theta _0}} \right){{\mathbf{R}}_X}{\mathbf{a}}\left( {{\theta _0}} \right) \le {{\mathbf{a}}^H}\left( {{\theta _k}} \right){{\mathbf{R}}_X}{\mathbf{a}}\left( {{\theta _k}} \right),\nonumber \\
  &\;\;\;\;\;\;\;\;\;\;\;\;\;\;\;\;\;\;\;\;\;\;\;\;\;\;\;\;\;\;\;\;\;\;\;\;\;\;\;\;\;\;\;\;\;\;\;\;\;\;\;\;\;\;\;\;\;\;\;\;\;\;\;\;\;\forall {\theta _k} \in \Phi  \hfill \\
  &\;\;\;\;\;\;\;{\text{SIN}}{{\text{R}}_i} \ge {\gamma _b}, \forall i, \hfill \\
  &\;\;\;\;\;\;\;\text{tr}({{\mathbf{R}}_{X}}) = {P_0},  \hfill \\
  &\;\;\;\;\;\;\;{{\mathbf{W}}_i} = {\mathbf{W}}_i^H, \; {{\mathbf{W}}_i} \succeq 0, \forall i, \hfill \\
  &\;\;\;\;\;\;\;{\text{rank}}\left( {{{\mathbf{W}}_i}} \right) = 1,\forall i, \hfill \\
  &\;\;\;\;\;\;\;{{\mathbf{R}}_N} = {\mathbf{R}}_N^H,\; {{\mathbf{R}}_N} \succeq 0,
\end{align}
\end{subequations}
where ${{\theta _0}}$ is the main-beam location, ${\Omega}$ denotes the sidelobe region of interest, ${\Phi}$ denotes the wide main-beam region, ${\gamma _s}$ is the bound of the sidelobe power.
\\\indent Likewise, recall the problem (13), SDR technique is adopted by neglecting rank-1 constraint in (17h). To solve the above sum-of-ratio problem, according to \cite{8314727}, we equivalently recast transform the minimization problem as
\begin{subequations}\label{eq18}
\begin{align}
  &\mathop {\max }\limits_{{{\mathbf{W}}_i},{{\mathbf{R}}_X}} \sum\limits_{{\theta _m} \in \Phi } {\frac{{{{\left| \alpha  \right|}^2}{{\mathbf{a}}^H}\left( {{\theta _m}} \right){{\mathbf{R}}_N}{\mathbf{a}}\left( {{\theta _m}} \right) + {\sigma ^2}}}{{{{\left| \alpha  \right|}^2}{{\mathbf{a}}^H}\left( {{\theta _m}} \right)\sum\nolimits_{i = 1}^K {{{\mathbf{W}}_i}} {\mathbf{a}}\left( {{\theta _m}} \right)}}}   \hfill \\
  &s.t.\;\;{{\mathbf{a}}^H}\left( {{\theta _0}} \right){{\mathbf{R}}_X}{\mathbf{a}}\left( {{\theta _0}} \right) - {{\mathbf{a}}^H}\left( {{\theta _m}} \right){{\mathbf{R}}_X}{\mathbf{a}}\left( {{\theta _m}} \right) \ge {\gamma _s}, \nonumber \\
  &\;\;\;\;\;\;\;\;\;\;\;\;\;\;\;\;\;\;\;\;\;\;\;\;\;\;\;\;\;\;\;\;\;\;\;\;\;\;\;\;\;\;\;\;\;\;\;\;\;\;\;\;\;\;\;\;\;\;\;\;\;\;\;\;\;\forall {\theta _m} \in \Omega  \hfill \\
  &\;\;\;\;\;\;\;{{\mathbf{a}}^H}\left( {{\theta _k}} \right){{\mathbf{R}}_X}{\mathbf{a}}\left( {{\theta _k}} \right) \le \left( {1 + \alpha } \right){{\mathbf{a}}^H}\left( {{\theta _0}} \right){{\mathbf{R}}_X}{\mathbf{a}}\left( {{\theta _0}} \right),\nonumber \\
  &\;\;\;\;\;\;\;\;\;\;\;\;\;\;\;\;\;\;\;\;\;\;\;\;\;\;\;\;\;\;\;\;\;\;\;\;\;\;\;\;\;\;\;\;\;\;\;\;\;\;\;\;\;\;\;\;\;\;\;\;\;\;\;\;\;\forall {\theta _k} \in \Phi  \hfill \\
  &\;\;\;\;\;\;\;\left( {1 - \alpha } \right){{\mathbf{a}}^H}\left( {{\theta _0}} \right){{\mathbf{R}}_X}{\mathbf{a}}\left( {{\theta _0}} \right) \le {{\mathbf{a}}^H}\left( {{\theta _k}} \right){{\mathbf{R}}_X}{\mathbf{a}}\left( {{\theta _k}} \right),\nonumber \\
  &\;\;\;\;\;\;\;\;\;\;\;\;\;\;\;\;\;\;\;\;\;\;\;\;\;\;\;\;\;\;\;\;\;\;\;\;\;\;\;\;\;\;\;\;\;\;\;\;\;\;\;\;\;\;\;\;\;\;\;\;\;\;\;\;\;\forall {\theta _k} \in \Phi  \hfill \\
  &\;\;\;\;\;\;\;{\text{SIN}}{{\text{R}}_i} \ge {\gamma _b}, \forall i, \hfill \\
  &\;\;\;\;\;\;\;\text{tr}({{\mathbf{R}}_{X}}) = {P_0},  \hfill \\
  &\;\;\;\;\;\;\;{{\mathbf{W}}_i} = {\mathbf{W}}_i^H, \; {{\mathbf{W}}_i} \succeq 0, \forall i, \hfill \\
  &\;\;\;\;\;\;\;{{\mathbf{R}}_N} = {\mathbf{R}}_N^H,\; {{\mathbf{R}}_N} \succeq 0.
\end{align}
\end{subequations}
It is noted that problem (18) is still non-convex. The approach to solve this sum-of-ratio FP problem is described in the following.
\subsection{Efficient Solver}
To present the solution to problem (18), we firstly refer to \cite{8314727} and denote
\begin{equation*}
\begin{aligned}
&A\left( {{\theta _m}} \right) = {\left| \alpha  \right|^2}{{\mathbf{a}}^H}\left( {{\theta _m}} \right){{\mathbf{R}}_N}{\mathbf{a}}\left( {{\theta _m}} \right) + {\sigma ^2} \hfill \\
&B\left( {{\theta _m}} \right) = {\left| \alpha  \right|^2}{{\mathbf{a}}^H}\left( {{\theta _m}} \right)\sum\nolimits_{i = 1}^K {{{\mathbf{W}}_i}} {\mathbf{a}}\left( {{\theta _m}} \right)
\end{aligned}
\end{equation*}
One step further, the sum-of-ratio problem is equivalent to the following optimization problem, which can be rewritten in the form
\begin{subequations}\label{eq19}
\begin{align}
    &\mathop {\max }\limits_{{{\mathbf{W}}_i},{{\mathbf{R}}_N},{\mathbf{y}}} \sum\limits_{{\theta _m} \in \Phi } {\left( {2{y_m}\sqrt {A\left( {{\theta _m}} \right)}  - y_m^2B\left( {{\theta _m}} \right)} \right)}  , \hfill \\
    &s.t.\;\;{{\mathbf{a}}^H}\left( {{\theta _0}} \right){{\mathbf{R}}_X}{\mathbf{a}}\left( {{\theta _0}} \right) - {{\mathbf{a}}^H}\left( {{\theta _m}} \right){{\mathbf{R}}_X}{\mathbf{a}}\left( {{\theta _m}} \right) \ge {\gamma _s}, \nonumber \\
    &\;\;\;\;\;\;\;\;\;\;\;\;\;\;\;\;\;\;\;\;\;\;\;\;\;\;\;\;\;\;\;\;\;\;\;\;\;\;\;\;\;\;\;\;\;\;\;\;\;\;\;\;\;\;\;\;\;\;\;\;\;\;\;\;\;\forall {\theta _m} \in \Omega  \hfill \\
    &\;\;\;\;\;\;\;{{\mathbf{a}}^H}\left( {{\theta _k}} \right){{\mathbf{R}}_X}{\mathbf{a}}\left( {{\theta _k}} \right) \le \left( {1 + \alpha } \right){{\mathbf{a}}^H}\left( {{\theta _0}} \right){{\mathbf{R}}_X}{\mathbf{a}}\left( {{\theta _0}} \right), \nonumber \\
    &\;\;\;\;\;\;\;\;\;\;\;\;\;\;\;\;\;\;\;\;\;\;\;\;\;\;\;\;\;\;\;\;\;\;\;\;\;\;\;\;\;\;\;\;\;\;\;\;\;\;\;\;\;\;\;\;\;\;\;\;\;\;\;\;\;\forall {\theta _k} \in \Phi  \hfill \\
    &\;\;\;\;\;\;\;\left( {1 - \alpha } \right){{\mathbf{a}}^H}\left( {{\theta _0}} \right){{\mathbf{R}}_X}{\mathbf{a}}\left( {{\theta _0}} \right) \le {{\mathbf{a}}^H}\left( {{\theta _k}} \right){{\mathbf{R}}_X}{\mathbf{a}}\left( {{\theta _k}} \right),\nonumber \\
    &\;\;\;\;\;\;\;\;\;\;\;\;\;\;\;\;\;\;\;\;\;\;\;\;\;\;\;\;\;\;\;\;\;\;\;\;\;\;\;\;\;\;\;\;\;\;\;\;\;\;\;\;\;\;\;\;\;\;\;\;\;\;\;\;\;\forall {\theta _k} \in \Phi  \hfill \\
    &\;\;\;\;\;\;\;{\text{SIN}}{{\text{R}}_i} \ge {\gamma _b}, \forall i, \hfill \\
    &\;\;\;\;\;\;\;\text{tr}({{\mathbf{R}}_{X}}) = {P_0}, \hfill \\
    &\;\;\;\;\;\;\;{{\mathbf{W}}_i} = {\mathbf{W}}_i^H, \; {{\mathbf{W}}_i} \succeq 0, \forall i, \hfill \\
    &\;\;\;\;\;\;\;{{\mathbf{R}}_N} = {\mathbf{R}}_N^H,\; {{\mathbf{R}}_N} \succeq 0,
\end{align}
\end{subequations}
where ${\mathbf{y}}$ denotes a collection of variables $\left\{ {{y_1}, \cdots ,{y_M}} \right\}$. The optimal ${y_m}$ can be obtained in the following closed form when $\theta_m$ is fixed
\begin{equation}\label{eq20}
    y_m^* = \frac{{\sqrt {A\left( {{\theta _m}} \right)} }}{{B\left( {{\theta _m}} \right)}}.
\end{equation}
To this end, the problem (19) can be solved by the SDR technique. Then, eigenvalue decomposition or Gaussian randomization is required to get the approximated solution. For clarity, the above procedure is summarized in Algorithm 2.
\renewcommand{\algorithmicrequire}{\textbf{Input:}}
\renewcommand{\algorithmicensure}{\textbf{Output:}}
\begin{algorithm}
\caption{Algorithm for solving sum-of-ratio problem (19)}
\label{alg:2}
\begin{algorithmic}
    \REQUIRE ${\mathbf{H}},{\mathbf{a}}\left( \theta  \right)$ or ${\mathbf{a}}\left( \theta_m  \right)$, ${\sigma ^2},\alpha ,{\gamma _b},{P_0},ite{r_{max}} \ge 2$, $\Delta\theta$.
    \ENSURE ${\mathbf{W}}_i^{\left( {iter} \right)},{\mathbf{R}}_N^{\left( {iter} \right)},i = 1, \cdots ,K$.
    \STATE 1. Compute ${{\mathbf{R}}_d}$. Reformulate problem (17) by (19). Set the iteration threshold $\varepsilon > 0$.
    \WHILE {$iter \le ite{r_{max}}$ and $\left\| {{{\mathbf{y}}^{iter + 1}} - {{\mathbf{y}}^{iter}}} \right\| \ge \varepsilon $}
    \STATE 2. Solve the new convex optimization problem.
    \STATE 3. Update ${\mathbf{y}}$ by (20).
    \STATE 4. Get updated ${{\mathbf{W}}_i}, \forall i,$ and ${{\mathbf{R}}_N}$ by solving (19) using SDR.
    \STATE 5. $ iter = iter + 1$.
    \ENDWHILE
\STATE 6. Obtain approximate solutions by eigenvalue decomposition or Gaussian randomization.
\end{algorithmic}
\end{algorithm}
\subsection{Complexity Analysis}
We end this section by computing the complexity of solving problem (19). It is noted that all the constraints can be considered as LMIs in optimization problem (19). We denote $\Phi_0=card(\Phi)$ and $\Omega_0=card(\Omega)$ as the cardinality of $\Phi$ and $\Omega$, respectively. Likely, eigenvalue decomposition operation is required as well, with the cost of $O\left( {\left( {K + 1} \right){N^3}} \right)$. Thus, referring to \cite{wang2014outage}, we give the computational complexity in Table I, which can be simplified as $\mathcal{O}\left( {3\sqrt 2 {N_{iter}}\ln \left( {{1 \mathord{\left/
 {\vphantom {1 \epsilon }} \right.
 \kern-\nulldelimiterspace} \epsilon }} \right){K^{3.5}}{N^{6.5}}} \right)+\mathcal{O}\left( {\left( {K + 1} \right){N^3}} \right)$ by reserving the highest order.
\section{Robust Beamforming With Imperfect CSI And Target Direction Uncertainty}
In this section, based on the models presented in the previous sections, we consider the case that perfect channel information is not available at the base station. By relying on the method of robust optimization, we generalize an optimization problem to obtain beamforming design that is robust to the channel uncertainty, which is bounded in a spherical region. Meanwhile, to guarantee the generality, we minimize the worst-case SINR received at the target in the angular interval of possible location of potential eavesdropper.
\subsection{Problem Formulation}
 According to \cite{6884062}, an additive channel error model of $i$-th downlink user can be formulated as ${{\mathbf{h}}_i} = {{\mathbf{\tilde h}}_i} + {{\mathbf{e}}_i}$, where ${{\mathbf{\tilde h}}_i}$ is the estimated channel information known at Alice, and ${{\mathbf{e}}_i}$ denotes the channel uncertainty within the spherical region ${\Im _i} = \{ {{\mathbf{e}}_i}|{\left\| {{{\mathbf{e}}_i}} \right\|^2} \le {\mathbf{\mu }}_i^2\}$. Following the well-known S-procedure , $\forall {\mathbf{e}}_i^H{{\mathbf{e}}_i} \le \mu _i^2$ , the constraint that guarantees the worst-case SINR of legitimates users can be reformulated as
\begin{equation}\label{eq21}
\begin{split}
    &{\left( {{{{\mathbf{\tilde h}}}_i} + {{\mathbf{e}}_i}} \right)^H}\left( {{{\mathbf{W}}_i} - {\gamma _b}\sum\limits_{k = 1,k \ne i}^K {{{\mathbf{W}}_k}}  - {\gamma _b}{{\mathbf{R}}_N}} \right)\left( {{{{\mathbf{\tilde h}}}_i} + {{\mathbf{e}}_i}}  \right) \hfill\\ &- {\gamma _b}{\sigma ^2} \ge 0,\forall i.
\end{split}
\end{equation}
Then, we minimize the possible maximum Eve SINR in the main-beam region of interest, which yields the following robust optimization problem
\begin{subequations}\label{eq22}
\begin{align}
    &\mathop {\min }\limits_{{{\mathbf{W}}_i},{{\mathbf{R}}_N},{t_i}} \mathop {\max }\limits_{{\theta _m} \in \Phi } \frac{{{{\left| \alpha  \right|}^2}{{\mathbf{a}}^H}\left( {{\theta _m}} \right)\sum\nolimits_{i = 1}^K {{{\mathbf{W}}_i}} {\mathbf{a}}\left( {{\theta _m}} \right)}}{{{{\left| \alpha  \right|}^2}{{\mathbf{a}}^H}\left( {{\theta _m}} \right){{\mathbf{R}}_N}{\mathbf{a}}\left( {{\theta _m}} \right) + {\sigma ^2}}} \hfill \\
    &s.t.\left( {\begin{array}{*{20}{c}}
    {{\mathbf{\tilde h}}_i^T{{\mathbf{Y}}_i}{\mathbf{\tilde h}}_i^* - {\gamma _b}{\sigma ^2} - {t_i}\mu _i^2}&{{\mathbf{\tilde h}}_i^T{{\mathbf{Y}}_i}} \\
    {{{\mathbf{Y}}_i}{\mathbf{\tilde h}}_i^*}&{{{\mathbf{Y}}_i} + {t_i}{{\mathbf{I}}_N}}
     \end{array}} \right) \succeq 0,\forall i, \hfill \\
     &\;\;\;\;{{\mathbf{Y}}_i}: = {{\mathbf{W}}_i} - {\gamma _b}\left( {\sum\limits_{k \ne i} {{{\mathbf{W}}_k}} } \right) - {\gamma _b}{{\mathbf{R}}_N} \nonumber \hfill \\
     &\;\;\;\;{{\mathbf{a}}^H}\left( {{\theta _0}} \right){{\mathbf{R}}_X}{\mathbf{a}}\left( {{\theta _0}} \right) - {{\mathbf{a}}^H}\left( {{\theta _m}} \right){{\mathbf{R}}_X}{\mathbf{a}}\left( {{\theta _m}} \right) \geqslant {\gamma _s}, \nonumber \\
    &\;\;\;\;\;\;\;\;\;\;\;\;\;\;\;\;\;\;\;\;\;\;\;\;\;\;\;\;\;\;\;\;\;\;\;\;\;\;\;\;\;\;\;\;\;\;\;\;\;\;\;\;\;\;\;\;\;\;\;\;\;\;\;\;\;\forall {\theta _m} \in \Omega  \hfill \\
    &\;\;\;\;{{\mathbf{a}}^H}\left( {{\theta _k}} \right){{\mathbf{R}}_X}{\mathbf{a}}\left( {{\theta _k}} \right) \leqslant \left( {1 + \alpha } \right){{\mathbf{a}}^H}\left( {{\theta _0}} \right){{\mathbf{R}}_X}{\mathbf{a}}\left( {{\theta _0}} \right), \nonumber \hfill \\
    &\;\;\;\;\;\;\;\;\;\;\;\;\;\;\;\;\;\;\;\;\;\;\;\;\;\;\;\;\;\;\;\;\;\;\;\;\;\;\;\;\;\;\;\;\;\;\;\;\;\;\;\;\;\;\;\;\;\;\;\;\;\;\;\;\;\forall {\theta _k} \in \Phi  \hfill \\
    &\;\;\;\;\left( {1 - \alpha } \right){{\mathbf{a}}^H}\left( {{\theta _0}} \right){{\mathbf{R}}_X}{\mathbf{a}}\left( {{\theta _0}} \right) \leqslant {{\mathbf{a}}^H}\left( {{\theta _k}} \right){{\mathbf{R}}_X}{\mathbf{a}}\left( {{\theta _k}} \right),\nonumber \hfill \\
    &\;\;\;\;\;\;\;\;\;\;\;\;\;\;\;\;\;\;\;\;\;\;\;\;\;\;\;\;\;\;\;\;\;\;\;\;\;\;\;\;\;\;\;\;\;\;\;\;\;\;\;\;\;\;\;\;\;\;\;\;\;\;\;\;\;\forall {\theta _k} \in \Phi  \hfill \\
    &\;\;\;\;\text{tr}({{\mathbf{R}}_{X}}) = {P_0},\hfill \\
    &\;\;\;\;{t_i} \geqslant 0,\forall i, \hfill \\
    &\;\;\;\;{{\mathbf{W}}_i} = {\mathbf{W}}_i^H,{{\mathbf{W}}_i} \succeq 0, \forall i, \hfill \\
    &\;\;\;\;{\text{rank}}\left( {{{\mathbf{W}}_i}} \right) = 1,\forall i, \hfill \\
    &\;\;\;\;{{\mathbf{R}}_N} = {\mathbf{R}}_N^H,{{\mathbf{R}}_N} \succeq 0,
\end{align}
\end{subequations}
where $\Phi=\left[ {{\theta _0} - \Delta \theta ,{\theta _0} + \Delta \theta } \right]$ is the main-beam region of interest, $m = 1, \cdots ,M$. $M$ represents the number of detecting angles in the interval $\Phi$,  and finally ${\mathbf{t}} = \left[ {{t_1}, \cdots ,{t_K}} \right]$ is an auxiliary vector relying on the S-procedure.
\subsection{Efficient Solver}
To solve problem (22), the SDR approach is adopted again by dropping the rank-1 constraint in (22i). Moreover, the objective function (22a) can be transformed to a max-min problem initially which is given as
\begin{equation}\label{eq23}
    \mathop {\max }\limits_{{{\mathbf{W}}_i},{{\mathbf{R}}_N},{t_i}} \mathop {\min }\limits_{{\theta _m} \in \Phi } \frac{{{{\left| \alpha  \right|}^2}{{\mathbf{a}}^H}\left( {{\theta _m}} \right){{\mathbf{R}}_N}{\mathbf{a}}\left( {{\theta _m}} \right) + {\sigma ^2}}}{{{{\left| \alpha  \right|}^2}{{\mathbf{a}}^H}\left( {{\theta _m}} \right)\sum\nolimits_{i = 1}^K {{{\mathbf{W}}_i}} {\mathbf{a}}\left( {{\theta _m}} \right)}}.
\end{equation}
To verify this, we introduce a variable $z$ and define ${\mathbf{A}}\left( {{\theta _m}} \right) = {\mathbf{a}}\left( {{\theta _m}} \right){{\mathbf{a}}^H}\left( {{\theta _m}} \right)$. The objective function (23) can be rewritten as $\mathop {\max }\limits_{{{\mathbf{W}}_i},{{\mathbf{R}}_N},{t_i},z} z$
, which subjects to $z \le {{\left( {{\text{tr}}\left( {{\mathbf{A}}\left( {{\theta _m}} \right){{\mathbf{R}}_N}} \right) + {{{\sigma ^2}} \mathord{\left/
 {\vphantom {{{\sigma ^2}} {{{\left| \alpha  \right|}^2}}}} \right.
 \kern-\nulldelimiterspace} {{{\left| \alpha  \right|}^2}}}} \right)} \mathord{\left/
 {\vphantom {{\left( {{\text{tr}}\left( {{\mathbf{A}}\left( {{\theta _m}} \right){{\mathbf{R}}_N}} \right) + {{{\sigma ^2}} \mathord{\left/
 {\vphantom {{{\sigma ^2}} {{{\left| \alpha  \right|}^2}}}} \right.
 \kern-\nulldelimiterspace} {{{\left| \alpha  \right|}^2}}}} \right)} {{\text{tr}}\left( {{\mathbf{A}}\left( {{\theta _m}} \right)\sum\nolimits_{i = 1}^K {{{\mathbf{W}}_i}} } \right)}}} \right.
 \kern-\nulldelimiterspace} {{\text{tr}}\left( {{\mathbf{A}}\left( {{\theta _m}} \right)\sum\nolimits_{i = 1}^K {{{\mathbf{W}}_i}} } \right)}}$ and any other contraints in (19). Likewise, we denote
\begin{equation*}
\begin{aligned}
    &C\left( {{\theta _m}} \right) = { {{\text{tr}}\left( {{\mathbf{A}}\left( {{\theta _m}} \right){{\mathbf{R}}_N}} \right) + {{{\sigma ^2}} \mathord{\left/
    {\vphantom {{{\sigma ^2}} {{{\left| \alpha  \right|}^2}}}} \right.
    \kern-\nulldelimiterspace} {{{\left| \alpha  \right|}^2}}}} } \hfill \\
    &D\left( {{\theta _m}} \right) = \mathord{
    {\vphantom {{\left( {{\text{tr}}\left( {{\mathbf{A}}\left( {{\theta _m}} \right){{\mathbf{R}}_N}} \right) + {{{\sigma ^2}} \mathord{\left/
    {\vphantom {{{\sigma ^2}} {{{\left| \alpha  \right|}^2}}}} \right.
    \kern-\nulldelimiterspace} {{{\left| \alpha  \right|}^2}}}} \right)} {{\text{tr}}\left( {{\mathbf{A}}\left( {{\theta _m}} \right)\sum\nolimits_{i = 1}^K {{{\mathbf{W}}_i}} } \right)}}}
    \kern-\nulldelimiterspace} {{\text{tr}}\left( {{\mathbf{A}}\left( {{\theta _m}} \right)\sum\nolimits_{i = 1}^K {{{\mathbf{W}}_i}} } \right)}
\end{aligned}
\end{equation*}
 The aforementioned constraint is equivalent to
 \begin{equation*}
    z \le {\max _{{y_m}}}\left( {2{y_m}\sqrt {C\left( {{\theta _m}} \right)}  - y_m^2D\left( {{\theta _m}} \right)} \right),
 \end{equation*}
 which is a less-than-max inequality, so $\mathop {\max }\limits_{{y_m}} $ can be integrated into the objective. Consequently, problem (22) is reformulated as
\begin{subequations}\label{eq24}
\begin{align}
    &\mathop {\max }\limits_{{{\mathbf{W}}_i},{{\mathbf{R}}_N},{\mathbf{y}},{t_i},z} z, \\
    \begin{split}
    &s.t.\;\;\;2{y_m}\sqrt {C\left( {{\theta _m}} \right)} - y_m^2{D\left( {{\theta _m}} \right)} \ge z,{\theta _m} \in \Phi ,\forall m,\\
    \end{split}\\
    &\;\;\;\;\left( {\begin{array}{*{20}{c}}
    {{\mathbf{\tilde h}}_i^T{{\mathbf{Y}}_i}{\mathbf{\tilde h}}_i^* - {\gamma _b}{\sigma ^2} - {t_i}\mu _i^2}&{{\mathbf{\tilde h}}_i^T{{\mathbf{Y}}_i}} \\
    {{{\mathbf{Y}}_i}{\mathbf{\tilde h}}_i^*}&{{{\mathbf{Y}}_i} + {t_i}{{\mathbf{I}}_N}}
     \end{array}} \right) \succeq 0,\forall i, \hfill \\
     &\;\;\;\;{{\mathbf{Y}}_i}: = {{\mathbf{W}}_i} - {\gamma _b}\left( {\sum\limits_{k \ne i} {{{\mathbf{W}}_k}} } \right) - {\gamma _b}{{\mathbf{R}}_N} \nonumber \hfill \\
    &\;\;\;\;{{\mathbf{a}}^H}\left( {{\theta _0}} \right){{\mathbf{R}}_X}{\mathbf{a}}\left( {{\theta _0}} \right) - {{\mathbf{a}}^H}\left( {{\theta _m}} \right){{\mathbf{R}}_X}{\mathbf{a}}\left( {{\theta _m}} \right) \ge {\gamma _s}, \nonumber \\
    &\;\;\;\;\;\;\;\;\;\;\;\;\;\;\;\;\;\;\;\;\;\;\;\;\;\;\;\;\;\;\;\;\;\;\;\;\;\;\;\;\;\;\;\;\;\;\;\;\;\;\;\;\;\;\;\;\;\;\;\;\;\;\;\;\;\forall {\theta _m} \in \Omega  \hfill \\
    &\;\;\;\;{{\mathbf{a}}^H}\left( {{\theta _k}} \right){{\mathbf{R}}_X}{\mathbf{a}}\left( {{\theta _k}} \right) \le \left( {1 + \alpha } \right){{\mathbf{a}}^H}\left( {{\theta _0}} \right){{\mathbf{R}}_X}{\mathbf{a}}\left( {{\theta _0}} \right), \nonumber \hfill \\
    &\;\;\;\;\;\;\;\;\;\;\;\;\;\;\;\;\;\;\;\;\;\;\;\;\;\;\;\;\;\;\;\;\;\;\;\;\;\;\;\;\;\;\;\;\;\;\;\;\;\;\;\;\;\;\;\;\;\;\;\;\;\;\;\;\;\forall {\theta _k} \in \Phi  \hfill \\
    &\;\;\;\;\left( {1 - \alpha } \right){{\mathbf{a}}^H}\left( {{\theta _0}} \right){{\mathbf{R}}_X}{\mathbf{a}}\left( {{\theta _0}} \right) \le {{\mathbf{a}}^H}\left( {{\theta _k}} \right){{\mathbf{R}}_X}{\mathbf{a}}\left( {{\theta _k}} \right),\nonumber \hfill \\
    &\;\;\;\;\;\;\;\;\;\;\;\;\;\;\;\;\;\;\;\;\;\;\;\;\;\;\;\;\;\;\;\;\;\;\;\;\;\;\;\;\;\;\;\;\;\;\;\;\;\;\;\;\;\;\;\;\;\;\;\;\;\;\;\;\;\forall {\theta _k} \in \Phi  \hfill \\
    &\;\;\;\;\text{tr}({{\mathbf{R}}_{X}}) = {P_0},\hfill \\
    &\;\;\;\;{t_i} \ge 0,\forall i, \hfill \\
    &\;\;\;\;{{\mathbf{W}}_i} = {\mathbf{W}}_i^H,{{\mathbf{W}}_i} \succeq 0, \forall i, \hfill \\
    &\;\;\;\;{{\mathbf{R}}_N} = {\mathbf{R}}_N^H,{{\mathbf{R}}_N} \succeq 0,
\end{align}
\end{subequations}
where ${y_m}$ is an auxiliary variable, each ${y_m}$ corresponds to the radar detecting angles $\theta_m $ in the main-beam region of interest $\Phi $. We refer the rest variables to the definitions which we presented in the previous sections. Note that problem (24) is convex and can be readily tackled. Here, we define a collection of variables ${\mathbf{y}} = \left\{ {{y_1}, \cdots ,{y_M}} \right\}$. To solve this problem, we apply the quadratic transform and optimize the primal variables ${{\mathbf{W}}_i},{{\mathbf{R}}_N},{t_i}$ and the auxiliary variable collection ${\mathbf{y}}$ in an alternating manner. When the primal variables are obtained by initializing the collection ${\mathbf{y}}$, the optimal ${y_m}$ can be updated by
\begin{equation}\label{eq25}
    y_m^* = \frac{{\sqrt {C\left( {{\theta _m}} \right)} }}{{D\left( {{\theta _m}} \right)}}.
\end{equation}
To this end, eigenvalue decomposition or Gaussian randomization is required to obtain approximated solutions. For clarity, solution to problem (24) can be summarized as Algorithm 3.
\renewcommand{\algorithmicrequire}{\textbf{Input:}}
\renewcommand{\algorithmicensure}{\textbf{Output:}}
\begin{algorithm}
\caption{Method for Solving multiple-ratio FP problem (24)}
\label{alg:3}
\begin{algorithmic}
    \REQUIRE ${\mathbf{{\rm A}}}\left( {{\theta _m}} \right),{{{\mathbf{\tilde h}}}_i},{\sigma ^2},\alpha ,{\gamma _b},{\gamma _s},{P_0}$, CSI estimation error threshold ${\mu _i} > 0$, definite the main-beam width $\Phi$, iteration threshold $\varepsilon > 0 $, $ite{r_{max}} \geqslant 2$.
    \STATE {\bf{Initialization}}: Set initial values for ${{\mathbf{y}}^{\left( 0 \right)}},{{\mathbf{y}}^{\left( 1 \right)}}$, which $\left\| {{{\mathbf{y}}^{\left( 1 \right)}} - {{\mathbf{y}}^{\left( 0 \right)}}} \right\| > \varepsilon $.
    \WHILE{$iter \leqslant ite{r_{max}}$ and $\left\| {{{\mathbf{y}}^{\left( {iter + 1} \right)}} - {{\mathbf{y}}^{\left( {iter} \right)}}} \right\| \geqslant \varepsilon $}
        \STATE 1. Reformulate problem (19) by replacing the fractional objective function with the form in (22b).
        \STATE 2. Reconstruct the problem with variable $z$.
        \STATE 3. Solve the optimization problem, and then update ${\mathbf{y}}$ by (23).
        \STATE 4. Update the primal variables by (22), over ${{{\mathbf{R}}_N}}, {{{\mathbf{W}}_i}}, \forall i$ for fixed ${\mathbf{y}}$.
    \ENDWHILE
    \ENSURE ${{{\mathbf{R}}_N}}, {{{\mathbf{W}}_i}}, {{t_i},z}, \forall i$.
    \STATE 6. Obtain approximate solutions by eigenvalue decomposition or Gaussian randomization.
\end{algorithmic}
\end{algorithm}
\subsection{Complexity Analysis}
The complexity of Algorithm 3 is analyzed in this subsection. Similarly, $\Phi$ and $\Omega$ can be regarded as discrete domains. We denote $\Phi_0=card(\Phi)$ and $\Omega_0=card(\Omega)$ as the cardinality of $\Phi$ and $\Omega$, respectively. All the constraints in problem (24) are LMIs. Specifically, we notice that the problem is composed by $3\Phi_0+\Omega_0+K+1$ LMI constraints of size 1, $2K+2$ LMI constraints of size $N$, and $K$ LMI constraints of size $N+1$. Considering eigenvalue decomposition operation is required at the cost of $O\left( {\left( {K + 1} \right){N^3}} \right)$, it follows that the complexity is given in Table I. For simplicity, we reserve the highest order of computational complexity, which can be given as $\mathcal{O}\left( {4\sqrt 3 {N_{iter}}\ln \left( {{1 \mathord{\left/
 {\vphantom {1 \epsilon }} \right.
 \kern-\nulldelimiterspace} \epsilon }} \right){K^{3.5}}{N^{6.5}}} \right)+\mathcal{O}\left( {\left( {K + 1} \right){N^3}} \right)$.

\section{Robust Optimal Beamforming With Statistical CSI And Target Direction Uncertainty}
In this section, we consider the extension of the scenario in section V, where channel from Alice to Bobs is rapidly time-varying. As a result, the instantaneous CSI is difficult to be estimated \cite{6671453}. Note that the second-order channel statistics, which vary much more slowly, can be obtained by the BS through long-term feedback. Nevertheless, even when the statistical CSI is known at Alice, it always includes uncertainty. Herewith, we take the uncertainty matrix into consideration by employing additive errors to the channel covariance.
\subsection{Problem Formulation}
As the statistical CSI is known to BS instead of instantaneous CSI , we rewrite the SINR of the \emph{i}-th user as
\begin{equation}\label{eq26}
{\text{SIN}}{{\text{R}}_i} = \frac{{{\text{tr}}\left( {{{{\mathbf{\tilde R}}}_{{h_i}}}{{\mathbf{W}}_i}} \right)}}{{\sum\nolimits_{k \ne i,k = 1}^K {{\text{tr}}\left( {{{{\mathbf{\tilde R}}}_{{h_i}}}{{\mathbf{W}}_k}} \right)}  + {\text{tr}}\left( {{{{\mathbf{\tilde R}}}_{{h_i}}}{{\mathbf{R}}_N}} \right) + {\sigma ^2}}},
\end{equation}
where ${{\mathbf{\tilde R}}_{{h_i}}} = {\mathbf{E}}\left\{ {{\mathbf{h}}_i^*{\mathbf{h}}_i^T} \right\}$ denotes the \emph{i}-th user's downlink channel covariance matrix with uncertainty. Therefore, the true channel covariance matrix can be modeled as ${{\mathbf{R}}_{{h_i}}} = {{\mathbf{\tilde R}}_{{h_i}}} + {{\mathbf{\Delta }}_i},\forall i$, where ${{\mathbf{\Delta }}_i}, \forall i$ are the estimated error matrices. The Frobenius norm of the error matrix of $i$-th user is assumed to be upper-bounded by a known constant ${\delta _i}$, which can be expressed as $\left\| {{{\mathbf{\Delta }}_i}} \right\| \le {\delta _i}$. To this end, based on Lagrange dual function \cite{wajid2009robust,law2017optimal}, the constraint corresponding to QoS of $i$-th user can be formulated as
\begin{equation*}
\begin{aligned}
&- {\delta _i}\left\| {{{\mathbf{A}}_i} + {{\mathbf{Z}}_{\mathbf{i}}}} \right\| - {\text{tr}}\left( {{{\mathbf{R}}_{{h_i}}}\left( {{{\mathbf{Z}}_i} + {{\mathbf{A}}_i}} \right)} \right) - {\gamma _b}{\text{tr}}\left( {{{\mathbf{R}}_{{h_i}}}{{\mathbf{R}}_N}} \right) \\
&- {\gamma _b}{\sigma ^2} \ge 0 \hfill \\
&{{\mathbf{Z}}_i} = {\mathbf{Z}}_i^H,{{\mathbf{Z}}_i} \succeq 0,\forall i
\end{aligned}
\end{equation*}
where ${{\mathbf{A}}_i} = {\gamma _b}\sum\nolimits_{k = 1,k \ne i}^K {{{\mathbf{W}}_k} - {{\mathbf{W}}_i}} ,\forall i$. Recalling the optimization problem in Section V-A, likewise, the robust beamforming problem with erroneous statistical CSI is given as
\begin{subequations}\label{eq27}
\begin{align}
  &\mathop {\min }\limits_{{{\mathbf{W}}_i},{{\mathbf{R}}_N},{{\mathbf{Z}}_i}} \mathop {\max }\limits_{{\theta _m} \in \Phi } \frac{{{{\left| \alpha  \right|}^2}{{\mathbf{a}}^H}\left( {{\theta _m}} \right)\sum\nolimits_{i = 1}^K {{{\mathbf{W}}_i}} {\mathbf{a}}\left( {{\theta _m}} \right)}}{{{{\left| \alpha  \right|}^2}{{\mathbf{a}}^H}\left( {{\theta _m}} \right){{\mathbf{R}}_N}{\mathbf{a}}\left( {{\theta _m}} \right) + {\sigma ^2}}} \hfill \\
  \begin{split}
  &s.t. \;\;- {\delta _i}\left\| {{{\mathbf{A}}_i} + {{\mathbf{Z}}_{i}}} \right\| - {\text{tr}}\left( {{{\mathbf{R}}_{{h_i}}}\left( {{{\mathbf{Z}}_i} + {{\mathbf{A}}_i}} \right)} \right) - {\gamma _b}{\text{tr}}\left( {{{\mathbf{R}}_{{h_i}}}{{\mathbf{R}}_N}} \right) \\
  &\;\;\;\;\;\;\; -{\gamma _b}{\sigma ^2} \ge 0, \forall i, \hfill \\
  \end{split}\hfill \\
    &\;\;\;\;{{\mathbf{a}}^H}\left( {{\theta _0}} \right){{\mathbf{R}}_X}{\mathbf{a}}\left( {{\theta _0}} \right) - {{\mathbf{a}}^H}\left( {{\theta _m}} \right){{\mathbf{R}}_X}{\mathbf{a}}\left( {{\theta _m}} \right) \ge {\gamma _s}, \nonumber \\
    &\;\;\;\;\;\;\;\;\;\;\;\;\;\;\;\;\;\;\;\;\;\;\;\;\;\;\;\;\;\;\;\;\;\;\;\;\;\;\;\;\;\;\;\;\;\;\;\;\;\;\;\;\;\;\;\;\;\;\;\;\;\;\;\;\;\forall {\theta _m} \in \Omega  \hfill \\
    &\;\;\;\;{{\mathbf{a}}^H}\left( {{\theta _k}} \right){{\mathbf{R}}_X}{\mathbf{a}}\left( {{\theta _k}} \right) \le \left( {1 + \alpha } \right){{\mathbf{a}}^H}\left( {{\theta _0}} \right){{\mathbf{R}}_X}{\mathbf{a}}\left( {{\theta _0}} \right), \nonumber \hfill \\
    &\;\;\;\;\;\;\;\;\;\;\;\;\;\;\;\;\;\;\;\;\;\;\;\;\;\;\;\;\;\;\;\;\;\;\;\;\;\;\;\;\;\;\;\;\;\;\;\;\;\;\;\;\;\;\;\;\;\;\;\;\;\;\;\;\;\forall {\theta _k} \in \Phi  \hfill \\
    &\;\;\;\;\left( {1 - \alpha } \right){{\mathbf{a}}^H}\left( {{\theta _0}} \right){{\mathbf{R}}_X}{\mathbf{a}}\left( {{\theta _0}} \right) \le {{\mathbf{a}}^H}\left( {{\theta _k}} \right){{\mathbf{R}}_X}{\mathbf{a}}\left( {{\theta _k}} \right),\nonumber \hfill \\
    &\;\;\;\;\;\;\;\;\;\;\;\;\;\;\;\;\;\;\;\;\;\;\;\;\;\;\;\;\;\;\;\;\;\;\;\;\;\;\;\;\;\;\;\;\;\;\;\;\;\;\;\;\;\;\;\;\;\;\;\;\;\;\;\;\;\forall {\theta _k} \in \Phi  \hfill \\
  &\;\;\;\;\;\text{tr}({{\mathbf{R}}_{X}}) = {P_0}, \hfill \\
  &\;\;\;\;\;{{\mathbf{Z}}_i} = {\mathbf{Z}}_i^H,{{\mathbf{Z}}_i} \succeq 0, \forall i, \hfill \\
  &\;\;\;\;\;{{\mathbf{W}}_i} = {\mathbf{W}}_i^H,{{\mathbf{W}}_i} \succeq 0, \forall i, \hfill \\
  &\;\;\;\;\;{\text{rank}}\left( {{{\mathbf{W}}_i}} \right) = 1,\forall i, \hfill \\
  &\;\;\;\;\;{{\mathbf{R}}_N} = {\mathbf{R}}_N^H,{{\mathbf{R}}_N} \succeq 0,
\end{align}
\end{subequations}
 We note that the problem (27) can be solved with SDR approach by dropping the rank-one constraint in (27i). One step further, similar to (22), problem (27) can be reformulated in a similar way, given by
\begin{subequations}\label{eq28}
\begin{align}
  &\mathop {\max }\limits_{{{\mathbf{W}}_i},{{\mathbf{R}}_N},{{\mathbf{Z}}_i},z} z \hfill \\
  \begin{split}
  &s.t.\;\;\;2{y_m}\sqrt {C\left( {{\theta _m}} \right)} - y_m^2{D\left( {{\theta _m}} \right)} \ge z,{\theta _m} \in \Phi ,\forall m,\\
 \end{split} \\
 \begin{split}
  & \;\;\;\;\;- {\delta _i}\left\| {{{\mathbf{A}}_i} + {{\mathbf{Z}}_{i}}} \right\| - {\text{tr}}\left( {{{\mathbf{R}}_{{h_i}}}\left( {{{\mathbf{Z}}_i} + {{\mathbf{A}}_i}} \right)} \right) - {\gamma _b}{\text{tr}}\left( {{{\mathbf{R}}_{{h_i}}}{{\mathbf{R}}_N}} \right) \\
  &\;\;\;\;\;- {\gamma _b}{\sigma ^2} \ge 0,\forall i, \hfill \\
  \end{split} \\
  &\;\;\;\;{{\mathbf{a}}^H}\left( {{\theta _0}} \right){{\mathbf{R}}_X}{\mathbf{a}}\left( {{\theta _0}} \right) - {{\mathbf{a}}^H}\left( {{\theta _m}} \right){{\mathbf{R}}_X}{\mathbf{a}}\left( {{\theta _m}} \right) \ge {\gamma _s}, \nonumber \\
    &\;\;\;\;\;\;\;\;\;\;\;\;\;\;\;\;\;\;\;\;\;\;\;\;\;\;\;\;\;\;\;\;\;\;\;\;\;\;\;\;\;\;\;\;\;\;\;\;\;\;\;\;\;\;\;\;\;\;\;\;\;\;\;\;\;\forall {\theta _m} \in \Omega  \hfill \\
    &\;\;\;\;{{\mathbf{a}}^H}\left( {{\theta _k}} \right){{\mathbf{R}}_X}{\mathbf{a}}\left( {{\theta _k}} \right) \le \left( {1 + \alpha } \right){{\mathbf{a}}^H}\left( {{\theta _0}} \right){{\mathbf{R}}_X}{\mathbf{a}}\left( {{\theta _0}} \right), \nonumber \hfill \\
    &\;\;\;\;\;\;\;\;\;\;\;\;\;\;\;\;\;\;\;\;\;\;\;\;\;\;\;\;\;\;\;\;\;\;\;\;\;\;\;\;\;\;\;\;\;\;\;\;\;\;\;\;\;\;\;\;\;\;\;\;\;\;\;\;\;\forall {\theta _k} \in \Phi  \hfill \\
    &\;\;\;\;\left( {1 - \alpha } \right){{\mathbf{a}}^H}\left( {{\theta _0}} \right){{\mathbf{R}}_X}{\mathbf{a}}\left( {{\theta _0}} \right) \le {{\mathbf{a}}^H}\left( {{\theta _k}} \right){{\mathbf{R}}_X}{\mathbf{a}}\left( {{\theta _k}} \right),\nonumber \hfill \\
    &\;\;\;\;\;\;\;\;\;\;\;\;\;\;\;\;\;\;\;\;\;\;\;\;\;\;\;\;\;\;\;\;\;\;\;\;\;\;\;\;\;\;\;\;\;\;\;\;\;\;\;\;\;\;\;\;\;\;\;\;\;\;\;\;\;\forall {\theta _k} \in \Phi  \hfill \\
  &\;\;\;\;\;\text{tr}({{\mathbf{R}}_{X}}) = {P_0}, \hfill \\
  &\;\;\;\;\;{{\mathbf{Z}}_i} = {\mathbf{Z}}_i^H,{{\mathbf{Z}}_i} \succeq 0, \hfill \\
  &\;\;\;\;\;{{\mathbf{W}}_i} = {\mathbf{W}}_i^H,{{\mathbf{W}}_i} \succeq 0, \hfill \\
  &\;\;\;\;\;{{\mathbf{R}}_N} = {\mathbf{R}}_N^H,{{\mathbf{R}}_N} \succeq 0.
\end{align}
\end{subequations}
Note that problem (28) is a convex SDP problem and can be solved in polynomial time using interior-point algorithms \cite{wajid2009robust}. To this end, approximated solution can be obtained by eigenvalue decomposition
or Gaussian randomization.
\subsection{Complexity Analysis}
The complexity of problem (27) is given as follows. As is noted in problem (28), almost all the constrains are LMI except for the SOC constraint (28c). Likewise, we denote $\Phi_0=card(\Phi)$ and $\Omega_0=card(\Omega)$ as the cardinality of $\Phi$ and $\Omega$. Note that the problem is composed by $K$ SOC constraints of size 1, $\Omega_0+3\Phi_0+1$ LMI constraints of size 1, and $4K+2$ LMIs of size $N$. Accordingly, we compute the complexity as is shown in Table I, which can be simply demonstrated as $\mathcal{O}\left( {5\sqrt 2 {N_{iter}}\ln \left( {{1 \mathord{\left/
 {\vphantom {1 \varepsilon }} \right.
 \kern-\nulldelimiterspace} \epsilon }} \right){K^{3.5}}{N^{6.5}}} \right)+\mathcal{O}\left( {\left( {K + 1} \right){N^3}} \right)$.
which is the complexity of each iteration. Then, The calculated complexities of all the proposed optimizations are summarised in Table 1.

\begin{table*}[htbp]
  \caption{Complexity Analysis}
  \label{table 1}
  \centering
\begin{tabular}{{lcl}}
\toprule
                 & Complexity \\
\midrule
\tabincell{c}{Perfect CSI and \\ Precise Target Location}         & \tabincell{c}{$\mathcal{O}\left( {{N_{iter}}\ln \left( {{1 \mathord{\left/
 {\vphantom {1 \epsilon }} \right.
 \kern-\nulldelimiterspace} \epsilon }} \right)\sqrt {2N\left( {K + 1} \right) + K + 3}  \cdot K{N^2}\left( {\left( {K + 1} \right)\left( {K{N^2} + 1} \right) + 2{N^3}\left( {{K^2}N + KN + K + 1} \right)} \right)} \right)$
 \\$+\mathcal{O}\left( {{N_{iter}}\ln \left( {{1 \mathord{\left/
 {\vphantom {1 \epsilon }} \right.
 \kern-\nulldelimiterspace} \epsilon }} \right)\sqrt {2N\left( {K + 1} \right) + K + 3}  \cdot K{N^4}\left( {{K^2}{N^2} + 1} \right)} \right)$
 \\$+ \mathcal{O}\left( {\left( {K + 1} \right){N^3}} \right)$}      \\
\tabincell{c}{Perfect CSI and \\ Target Location Uncertainty }    & \tabincell{c}{ $\mathcal{O}\left( {{N_{iter}}\ln \left( {{1 \mathord{\left/
 {\vphantom {1 \epsilon }} \right.
 \kern-\nulldelimiterspace} \epsilon }} \right)\sqrt {2N\left( {K + 1} \right) + K + {\Omega _0} + 2{\Phi _0} + 1}  \cdot K{N^2}\left( {K{N^2} + 1} \right)\left( {K + {\Omega _0} + 2{\Phi _0} + 1} \right)} \right)$
 \\$+\mathcal{O}\left( {{N_{iter}}\ln \left( {{1 \mathord{\left/
 {\vphantom {1 \epsilon }} \right.
 \kern-\nulldelimiterspace} \epsilon }} \right)\sqrt {2N\left( {K + 1} \right) + K + {\Omega _0} + 2{\Phi _0} + 1}  \cdot K{N^2}\left( {2{N^3}\left( {{K^2}N + KN + K + 1} \right) + {K^2}{N^4}} \right)} \right)$
 \\$+ \mathcal{O}\left( {\left( {K + 1} \right){N^3}} \right)$ }     \\
\tabincell{c}{Imperfect CSI and \\Target Location Uncertainty}   & \tabincell{c}{
$\mathcal{O}\left( {{N_{iter}}\ln \left( {{1 \mathord{\left/
 {\vphantom {1 \epsilon }} \right.
 \kern-\nulldelimiterspace} \epsilon }} \right)\sqrt {3NK + 2\left( {K + N + {\Omega _0} + {\Phi _0}} \right) + 1}  \cdot K{N^2}\left( {K{N^2} + 1} \right)\left( {K + {\Omega _0} + 3{\Phi _0} + 1} \right)} \right)$\\$+\mathcal{O}\left( {{N_{iter}}\ln \left( {{1 \mathord{\left/
 {\vphantom {1 \epsilon }} \right.
 \kern-\nulldelimiterspace} \epsilon }} \right)\sqrt {3NK + 2\left( {K + N + {\Omega _0} + {\Phi _0}} \right) + 1}  \cdot 2K{N^5}\left( {K + 1} \right)\left( {KN + 1} \right)} \right)$\\$+\mathcal{O}\left( {{N_{iter}}\ln \left( {{1 \mathord{\left/
 {\vphantom {1 \epsilon }} \right.
 \kern-\nulldelimiterspace} \epsilon }} \right)\sqrt {3NK + 2\left( {K + N + {\Omega _0} + {\Phi _0}} \right) + 1}  \cdot K{N^2}\left( {K{{\left( {N + 1} \right)}^2}\left( {K{N^2} + N + 1} \right) + {K^2}{N^4}} \right)} \right)$\\$+ \mathcal{O}\left( {\left( {K + 1} \right){N^3}} \right)$}     \\
\tabincell{c}{Statistical CSI and \\Target Location Uncertainty} & \tabincell{c}{$\mathcal{O}\left( {{N_{iter}}\ln \left( {{1 \mathord{\left/
 {\vphantom {1 \epsilon }} \right.
 \kern-\nulldelimiterspace} \epsilon }} \right)\sqrt {2N\left( {2K + 1} \right) + 3{\Phi _0} + {\Omega _0} + 1}  \cdot K{N^2}\left( {\left( {K{N^2} + 1} \right)\left( {3{\Phi _0} + {\Omega _0} + 1} \right) + K} \right)} \right)$\\$+\mathcal{O}\left( {{N_{iter}}\ln \left( {{1 \mathord{\left/
 {\vphantom {1 \epsilon }} \right.
 \kern-\nulldelimiterspace} \epsilon }} \right)\sqrt {2N\left( {2K + 1} \right) + 3{\Phi _0} + {\Omega _0} + 1}  \cdot K{N^2}\left( {2{N^3}\left( {2K + 1} \right)\left( {KN + 1} \right) + {K^2}{N^4}} \right)} \right)$ \\$ + \mathcal{O}\left( {\left( {K + 1} \right){N^3}} \right)$}      \\
\bottomrule
\end{tabular}
\end{table*}
\begin{figure*}[htb]
\centering
\subfigure[]{
    \begin{minipage}[b]{0.3\textwidth}
    \includegraphics[width=2.4in,height=2in]{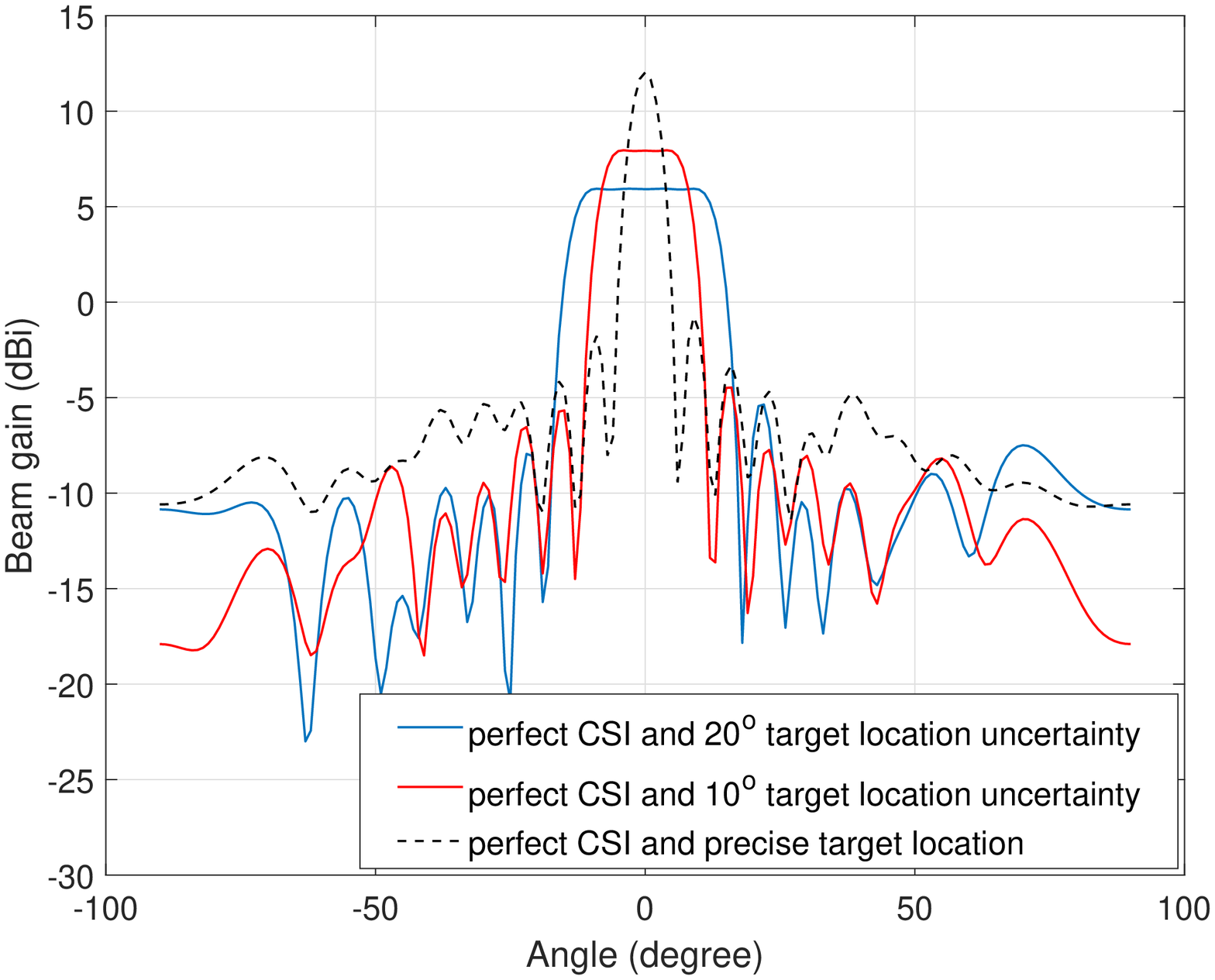}
    \end{minipage}
}
\subfigure[]{
    \begin{minipage}[b]{0.3\textwidth}
    \includegraphics[width=2.4in,height=2in]{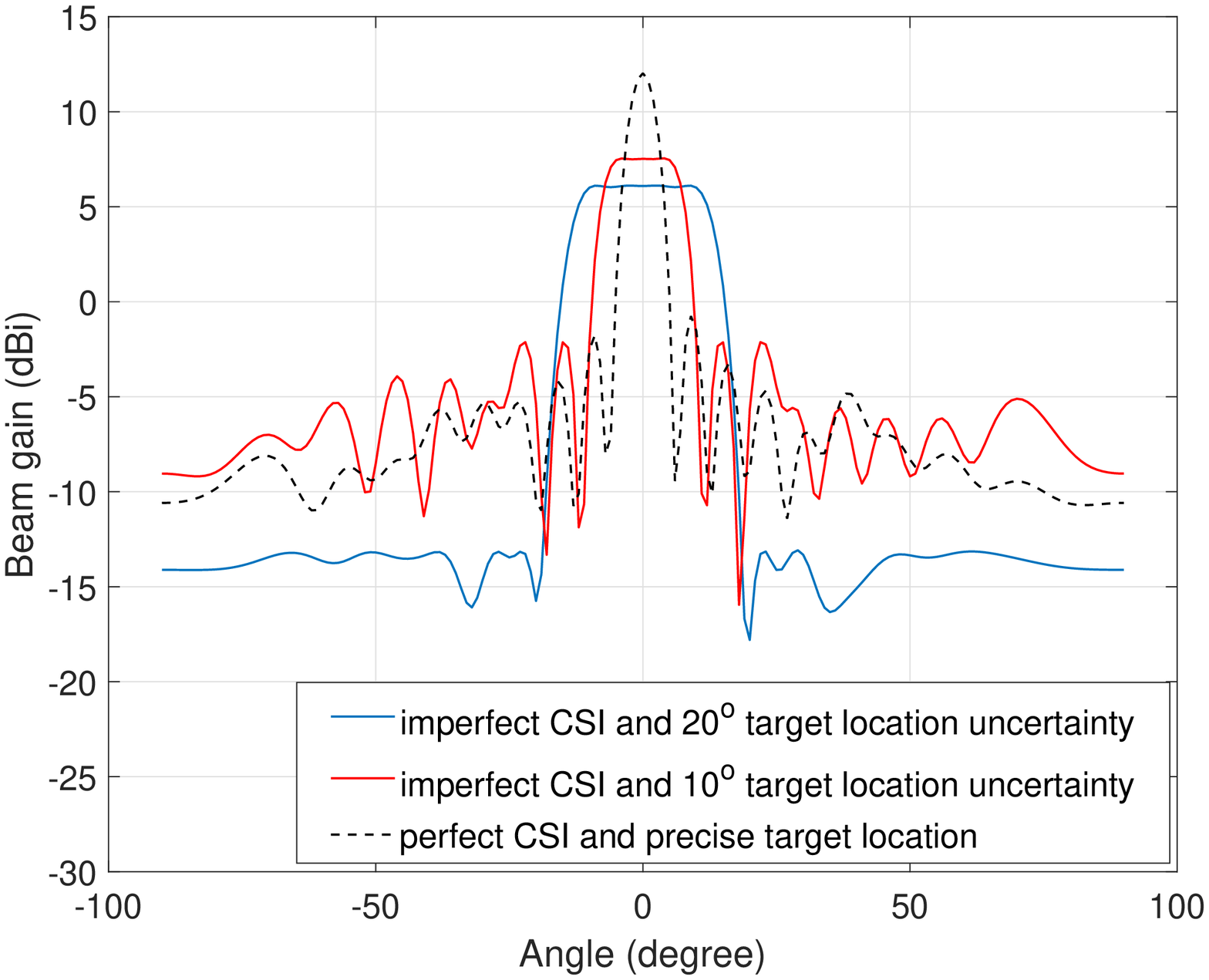}
    \end{minipage}
}
\subfigure[]{
    \begin{minipage}[b]{0.3\textwidth}
    \includegraphics[width=2.4in,height=2in]{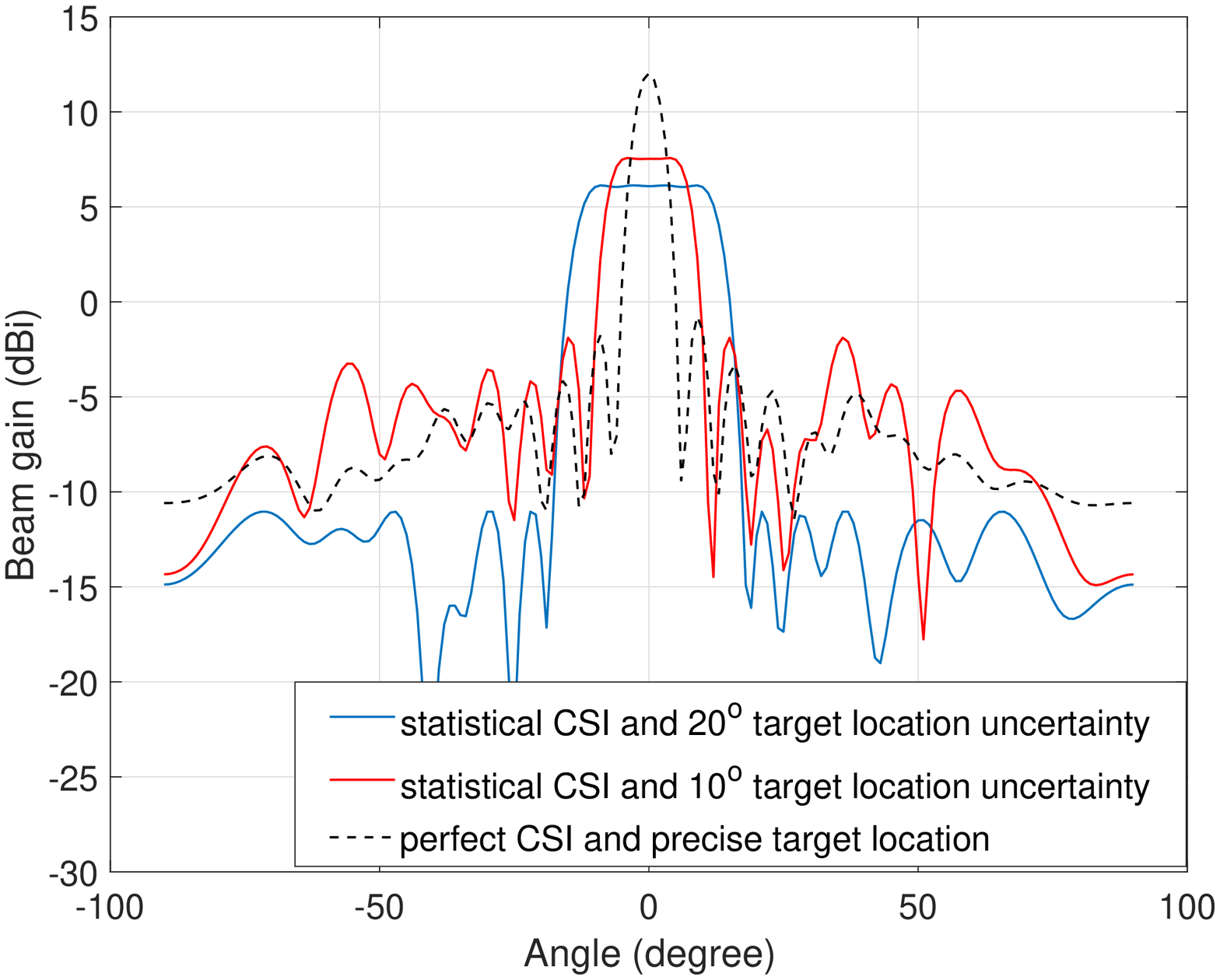}
    \end{minipage}
}
\caption{Beampatterns with various target direction uncertainty interval when (a) CSI is known, (b) CSI is imperfectly known and (c) statistical CSI is imperfectly known.} \label{fig.2}
\end{figure*}
\begin{figure}[H]
    \centering
    \includegraphics[width=\columnwidth]{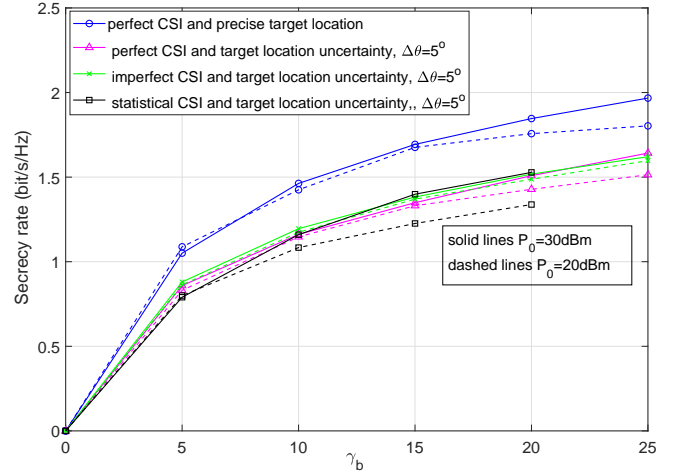}
    \captionsetup{font={footnotesize}}
    \caption{Achievable secrecy rate versus the threshold of SINR at legitimate users, with various transmission power budget, where solid and dashed lines represent power budget $P_0=30$dBm and $P_0=20$dBm respectively, $N=18, K=4, \Delta\theta=5^\circ$.}
    \label{fig.3}
\end{figure}
\section{Numerical Results}
To evaluate the proposed methods, numerical results based on Monte Carlo simulations are shown in this section to validate the effectiveness of the proposed beamforming method. Without loss of generality, each entry of channel matrix $\mathbf{H}$ is assumed to obey standard Complex Gaussian distribution, i.e. ${h_{i,j}} \sim \mathcal{C}\mathcal{N}\left( {0,1} \right)$. We assume that the DFRC base station employs a ULA with half-wavelength spacing between adjacent antennas. In the following simulations, the number of antennas is set as $N=18$ and the number of legitimate users is $K=4$. The constrained beamforming design problems in Section II-Section V are solved by the classic SDR technique using the CVX toolbox \cite{grant2008cvx}.

\subsection{Beam Gain And Secrecy Rate Analysis}
We first show the resultant radar beampattern in Fig. 2 with different angular interval of target location uncertainty, i.e. $\left[ { - {5^ \circ },{5^ \circ }} \right]$ and $\left[ { - {10^ \circ },{10^ \circ }} \right]$. The SINR threshold of each legitimate user is set as $\gamma_b=10\text{dB}$. The narrow beampattern when the target location is precisely known at the BS is set as a benchmark. It is found that the desired beampattern with wide main-beam is obtained by solving the proposed algorithms, which maintain the same power in the region of possible target location. Additionally, it is noted that with the expansion of location uncertainty angular interval, the power gain of main-beam reduces.
\\\indent The achievable secrecy rate in terms of increasing SINR threshold of each user is demonstrated in Fig. 3, where the power budget is set as $P_0=20\text{dBm}$ and $P_0=30\text{dBm}$ respectively. In this case, we set the sidelobe power threshold $\gamma_s=40\text{dB}$. Basically, in the $\text{SINR}_E$ minimization problem, the secrecy rate increases with the growth of $\gamma_b$. It is noteworthy that the system achieves higher secrecy rate when both the target location and CSI are precisely known. Besides, when we increase the power budget, the secrecy rate grows to some extent.
\begin{figure}[htb]
  \begin{minipage}{0.235\textwidth}
    \includegraphics[width=1.7in,height=2.7in]{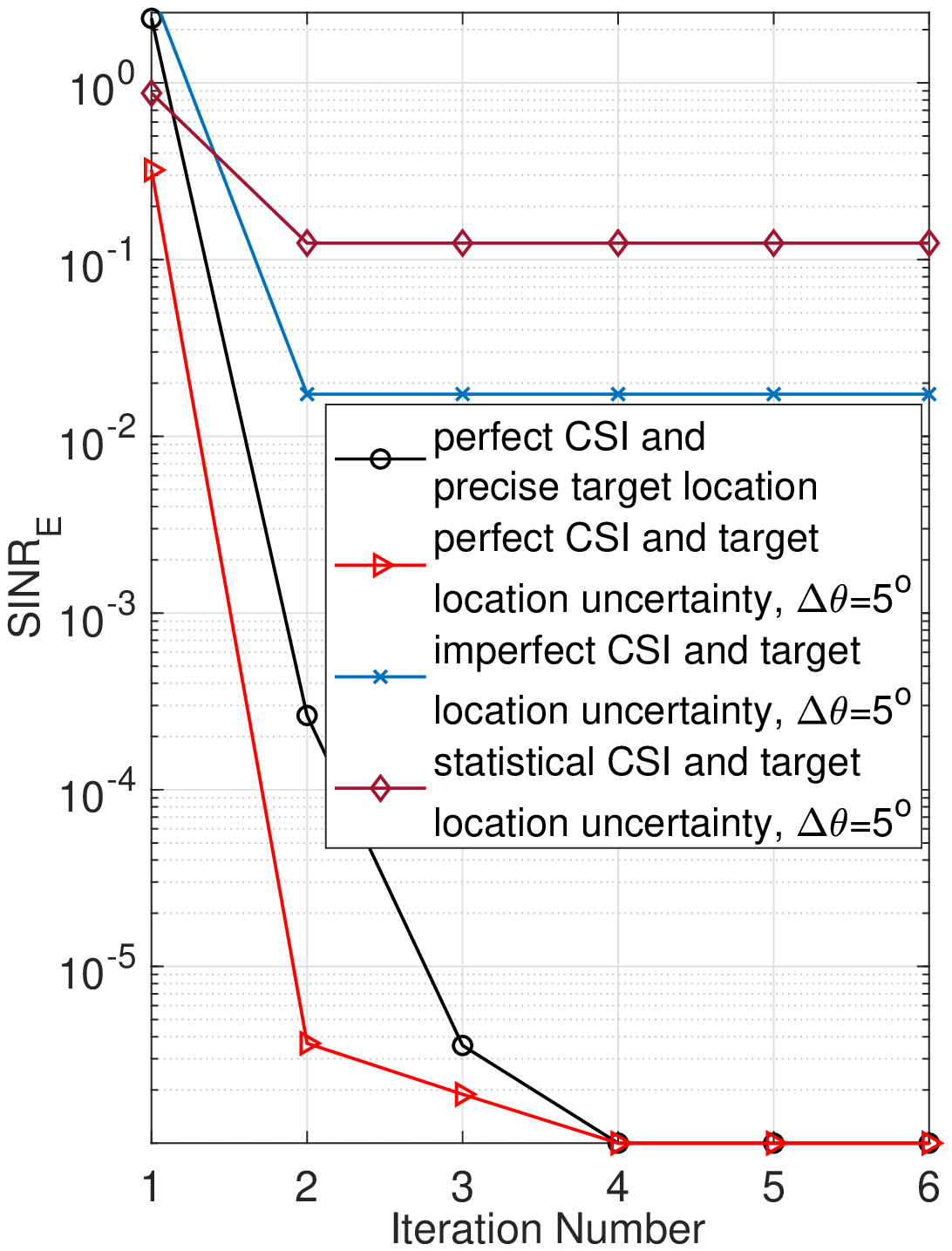}
    \captionsetup{font={footnotesize}}
  \end{minipage}
  \hfill
    \begin{minipage}{0.235\textwidth}
    \includegraphics[width=1.7in,height=2.7in]{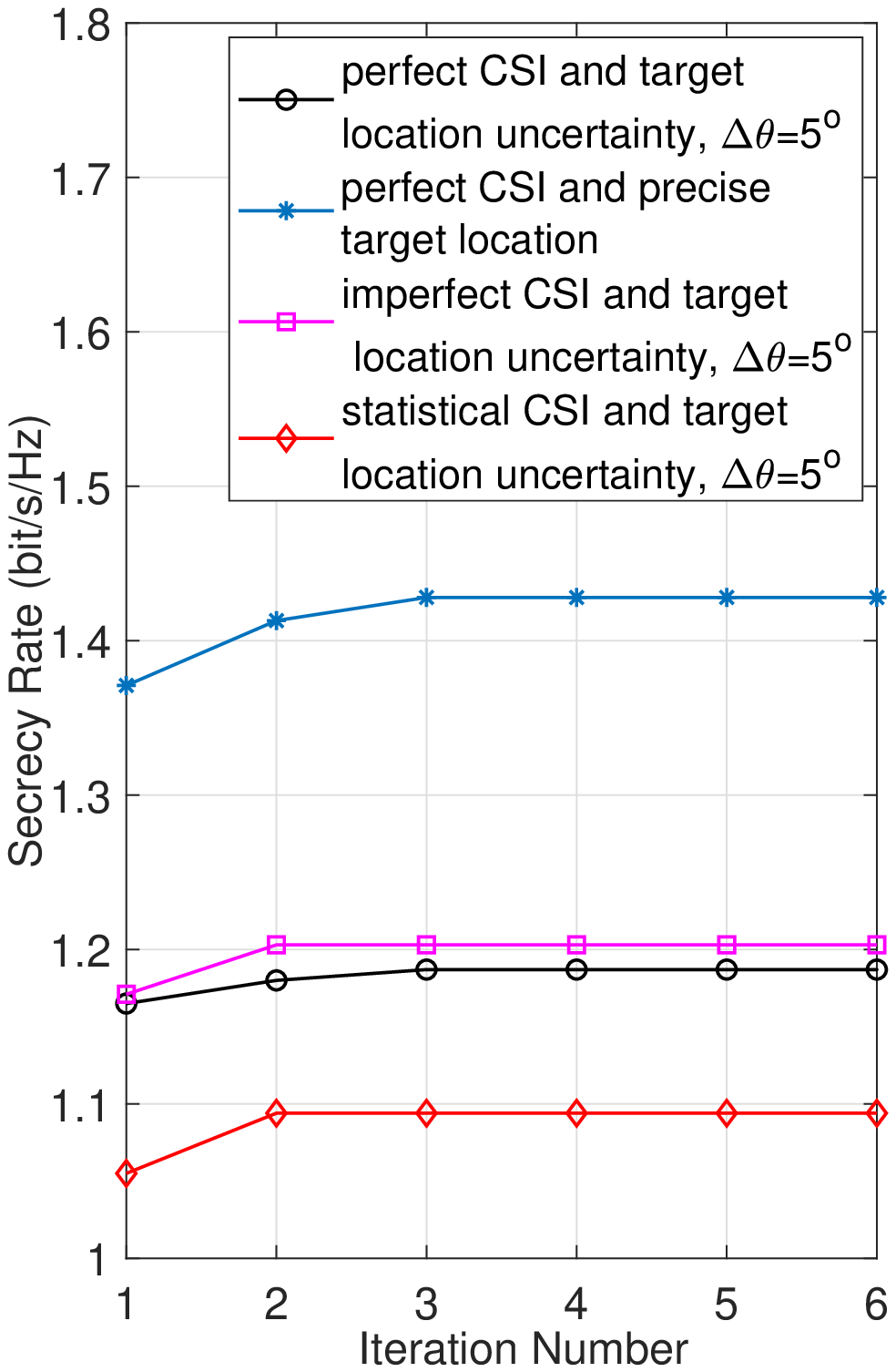}
    \captionsetup{font={footnotesize}}
  \end{minipage}
  \hfill
  \captionsetup{font={footnotesize}}
  \caption{Convergence of (a) SINR of Eve and (b) secrecy rate for the target SINR minimization algorithm, $N=18, K=4, P_0=30\text{dBm}, \gamma_b=10\text{dB}$.}
  \label{fig.4}
\end{figure}
\\\indent In Fig. 4, we evaluate the convergence of target SINR and secrecy rate. In these cases, the same system parameters are set as previous simulations. In Fig. 4(a), the SINR of the target is confirmed to convergent to a minimum. In robust beamforming design problems, the SINR of target decreases slightly with the increasing iteration number, which results in the slight growth of secrecy rate as is shown in Fig. 4(b).
\begin{figure}
    \centering
    \includegraphics[width=\columnwidth]{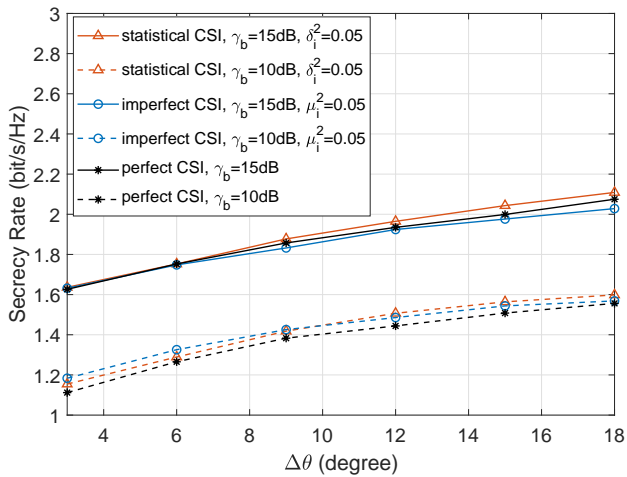}
    \captionsetup{font={footnotesize}}
    \caption{Secrecy rate with different angular intervals, $N=18, K=4, P_0=30\text{dBm}$, with $\gamma_b=10\text{dB}$ and $\gamma_b=15\text{dB}$, respectively.}
    \label{fig.5}
\end{figure}
\subsection{Trade-off Between The Performance Of Radar And Communication System}
In this subsection, we evaluate the performance trade-off between radar and communication system. Fig. 5 shows the secrecy rate performance with various angular intervals for $\gamma_b=10\text{dB}$ and $\gamma_b=15\text{dB}$. The main-beam power decreases when the target uncertainty increases, then the leaking information would get less, which improve the secrecy rate. As is demonstrated in Fig. 5, the secrecy rate increases with the growth of target uncertainty interval. Besides, with $5\text{dB}$ growth of legitimate user SINR threshold, the secrecy rate increases $0.5\text{bit/s/Hz}$ approximately.
\\\indent Fig. 6 demonstrates the secrecy rate performance versus the threshold of sidelobe with $P_0=30\text{dBm}, \Delta\theta=5^\circ$, which reveals the trade-off between the performance of radar and communication systems. In Algorithm 2, the power difference between main beam and sidelobe increases with the growth of $\gamma_s$, which results in the increasing possibility of information leaking. As the numerical result shown in Fig. 6, it is notable that the secrecy rate decreases with the growth of $\gamma_s$, especially the tendency gets obvious when $\gamma_s$ is greater than 30dB.
\begin{figure}
    \centering
    \includegraphics[width=\columnwidth]{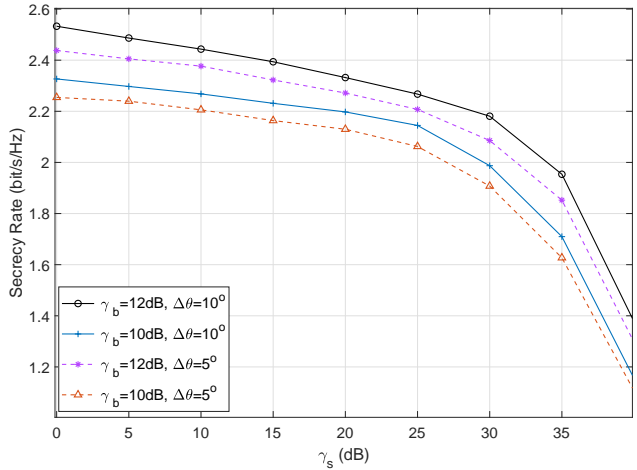}
    \captionsetup{font={footnotesize}}
    \caption{Achievable secrecy rate versus the sidelobe power with various SINR threshold of legitimate users for the Algorithm 2, $N=18, K=4, P_0=30\text{dBm}, \Delta\theta=5^\circ$.}
    \label{fig.6}
\end{figure}
\subsection{Robust Beamforming Performance}
As the norm of CSI error is bounded by a constant, the secrecy rate performance versus error bound is illustrated in Fig. 7, with different location uncertainty. With the growth of error bound, the achievable SINR at each legitimate user keeps being above the given threshold but not a constant according to constraints (24c) and (27b). We note that the achievable secrecy rate reduces after a certain value with the increasing error bound, because of the different changing rate between target SINR and user SINR corresponding to various error bounds in Fig. 7. Whereas, as is shown in Fig. 8, the secrecy rate keeps increasing with the growth of error bound. In addition, the robust beamforming designs achieve higher secrecy rate when the location uncertainty is limited in a larger interval.
\begin{figure}
    \centering
    \includegraphics[width=\columnwidth]{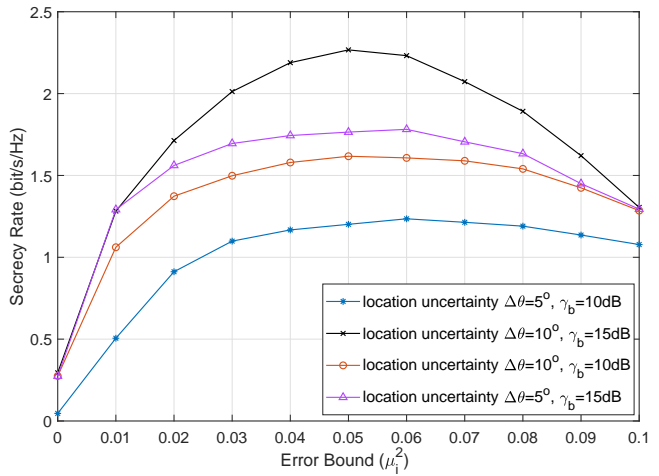}
    \captionsetup{font={footnotesize}}
    \caption{Achieved secrecy rate with different error bounds in the scenario of known imperfect CSI, $N=18, K=4, P_0=30\text{dBm}$.}
    \label{fig.7}
\end{figure}

\begin{figure}
    \centering
    \includegraphics[width=\columnwidth]{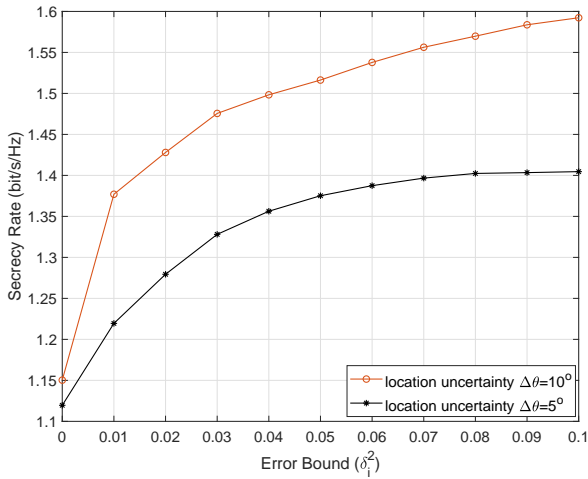}
    \captionsetup{font={footnotesize}}
    \caption{Achieved secrecy rate versus different error bounds when statistical CSI is imperfectly known, $N=18, K=4, P_0=30\text{dBm}, \gamma_b=10\text{dB}$.}
    \label{fig.8}
\end{figure}
\section{Conclusion}
In this paper, optimization based beamforming designs have been addressed for MIMO DFRC system, which aimed at ensuring the security of information transmission in case of leaking to targets by adding AN at the transmitter to confuse the potential eavesdropper. Specifically, we have minimized the SINR of the target which is regarded as the potential eavesdropper while keeping the each legitimate user's SINR above a certain constant to ensure the secrecy rate of the DFRC system. Throughout this paper, the optimization beamforming problem has been designed with perfect CSI and imperfect CSI, as well as with the accurate and inaccurate target location information.
\\\indent First of all, both precise location of target and perfect CSI have been assumed to be known at BS, which gained the highest secrecy rate according to the numerical results. When the target location was uncertain, the main-beam power has decreased with the growth of the uncertainty angular interval. Moreover, the secrecy rate versus different thresholds of sidelobe has been demonstrated, which revealed the trade-off between radar and communication system performance. Then, we have formulated target SINR minimization problem with imperfect instantaneous CSI and statistical CSI known to the base station respectively. As shown in the numerical results, the beamforming design has been feasible in both robust scenarios. Finally, simulation results have been presented to show the secrecy rate tendency effected by error bound with various target location uncertainty.


\ifCLASSOPTIONcaptionsoff
  \newpage
\fi



\bibliographystyle{IEEEtran}
\bibliography{IEEEabrv,CEP_REF}

\begin{thebibliography}{10}
\providecommand{\url}[1]{#1}
\csname url@samestyle\endcsname
\providecommand{\newblock}{\relax}
\providecommand{\bibinfo}[2]{#2}
\providecommand{\BIBentrySTDinterwordspacing}{\spaceskip=0pt\relax}
\providecommand{\BIBentryALTinterwordstretchfactor}{4}
\providecommand{\BIBentryALTinterwordspacing}{\spaceskip=\fontdimen2\font plus
\BIBentryALTinterwordstretchfactor\fontdimen3\font minus
  \fontdimen4\font\relax}
\providecommand{\BIBforeignlanguage}[2]{{%
\expandafter\ifx\csname l@#1\endcsname\relax
\typeout{** WARNING: IEEEtran.bst: No hyphenation pattern has been}%
\typeout{** loaded for the language `#1'. Using the pattern for}%
\typeout{** the default language instead.}%
\else
\language=\csname l@#1\endcsname
\fi
#2}}
\providecommand{\BIBdecl}{\relax}
\BIBdecl

\bibitem{oyediran2015spectrum}
D.~Oyediran, ``Spectrum sharing: Overview and challenges of small cells
  innovation in the proposed 3.5 {GHz} band.''\hskip 1em plus 0.5em minus
  0.4em\relax International Foundation for Telemetering, 2015.

\bibitem{li2016optimum}
B.~Li, A.~P. Petropulu, and W.~Trappe, ``Optimum co-design for spectrum sharing
  between matrix completion based {MIMO} radars and a {MIMO} communication
  system,'' \emph{IEEE Transactions on Signal Processing}, vol.~64, no.~17, pp.
  4562--4575, 2016.

\bibitem{1542627}
A.~{Ghasemi} and E.~S. {Sousa}, ``Collaborative spectrum sensing for
  opportunistic access in fading environments,'' in \emph{First IEEE
  International Symposium on New Frontiers in Dynamic Spectrum Access Networks,
  2005. DySPAN 2005.}, Nov 2005, pp. 131--136.

\bibitem{kim2015design}
C.~W. Kim, J.~Ryoo, and M.~M. Buddhikot, ``Design and implementation of an
  end-to-end architecture for 3.5 {GHz} shared spectrum,'' in \emph{2015 IEEE
  International Symposium on Dynamic Spectrum Access Networks (DySPAN)}.\hskip
  1em plus 0.5em minus 0.4em\relax IEEE, 2015, pp. 23--34.

\bibitem{staple2004end}
G.~Staple and K.~Werbach, ``The end of spectrum scarcity [spectrum allocation
  and utilization],'' \emph{IEEE spectrum}, vol.~41, no.~3, pp. 48--52, 2004.

\bibitem{sodagari2012projection}
S.~Sodagari, A.~Khawar, T.~C. Clancy, and R.~McGwier, ``A projection based
  approach for radar and telecommunication systems coexistence,'' in \emph{2012
  IEEE Global Communications Conference (GLOBECOM)}.\hskip 1em plus 0.5em minus
  0.4em\relax IEEE, 2012, pp. 5010--5014.

\bibitem{turlapaty2014joint}
A.~Turlapaty and Y.~Jin, ``A joint design of transmit waveforms for radar and
  communications systems in coexistence,'' in \emph{2014 IEEE Radar
  Conference}.\hskip 1em plus 0.5em minus 0.4em\relax IEEE, 2014, pp.
  0315--0319.

\bibitem{li2017joint}
B.~Li and A.~P. Petropulu, ``Joint transmit designs for coexistence of {MIMO}
  wireless communications and sparse sensing radars in clutter,'' \emph{IEEE
  Transactions on Aerospace and Electronic Systems}, vol.~53, no.~6, pp.
  2846--2864, 2017.

\bibitem{8288677}
F.~Liu, C.~Masouros, A.~Li, H.~Sun, and L.~Hanzo, ``{MU-MIMO} communications
  with {MIMO} radar: From co-existence to joint transmission,'' \emph{IEEE
  Transactions on Wireless Communications}, vol.~17, no.~4, pp. 2755--2770,
  April 2018.

\bibitem{liu2017robust}
F.~Liu, C.~Masouros, A.~Li, and T.~Ratnarajah, ``Robust {MIMO} beamforming for
  cellular and radar coexistence,'' \emph{IEEE Wireless Communications
  Letters}, vol.~6, no.~3, pp. 374--377, 2017.

\bibitem{liu2018mimo}
F.~Liu, C.~Masouros, A.~Li, T.~Ratnarajah, and J.~Zhou, ``Mimo radar and
  cellular coexistence: A power-efficient approach enabled by interference
  exploitation,'' \emph{IEEE Transactions on Signal Processing}, vol.~66,
  no.~14, pp. 3681--3695, 2018.

\bibitem{8386661}
F.~{Liu}, L.~{Zhou}, C.~{Masouros}, A.~{Li}, W.~{Luo}, and A.~{Petropulu},
  ``Toward dual-functional radar-communication systems: Optimal waveform
  design,'' \emph{IEEE Transactions on Signal Processing}, vol.~66, no.~16, pp.
  4264--4279, Aug 2018.

\bibitem{liu2018dual}
F.~Liu, L.~Zhou, C.~Masouros, A.~Lit, W.~Luo, and A.~Petropulu,
  ``Dual-functional cellular and radar transmission: Beyond coexistence,'' in
  \emph{2018 IEEE 19th International Workshop on Signal Processing Advances in
  Wireless Communications (SPAWC)}.\hskip 1em plus 0.5em minus 0.4em\relax
  IEEE, 2018, pp. 1--5.

\bibitem{zhou2018optimal}
L.~Zhou, F.~Liu, C.~Tian, C.~Masouros, A.~Li, W.~Jiang, and W.~Luo, ``Optimal
  waveform design for dual-functional {MIMO} radar-communication systems,'' in
  \emph{2018 IEEE/CIC International Conference on Communications in China
  (ICCC)}.\hskip 1em plus 0.5em minus 0.4em\relax IEEE, 2018, pp. 661--665.

\bibitem{hassanien2016dual}
A.~Hassanien, M.~G. Amin, Y.~D. Zhang, and F.~Ahmad, ``Dual-function
  radar-communications: Information embedding using sidelobe control and
  waveform diversity,'' \emph{IEEE Transactions on Signal Processing}, vol.~64,
  no.~8, pp. 2168--2181, 2016.

\bibitem{va2016millimeter}
V.~Va, T.~Shimizu, G.~Bansal, R.~W. Heath~Jr \emph{et~al.}, ``Millimeter wave
  vehicular communications: A survey,'' \emph{Foundations and
  Trends{\textregistered} in Networking}, vol.~10, no.~1, pp. 1--113, 2016.

\bibitem{lim2007real}
C.-H. Lim, Y.~Wan, B.-P. Ng, and C.-M.~S. See, ``A real-time indoor wifi
  localization system utilizing smart antennas,'' \emph{IEEE Transactions on
  Consumer Electronics}, vol.~53, no.~2, pp. 618--622, 2007.

\bibitem{dillard2003cyclic}
G.~M. Dillard, M.~Reuter, J.~Zeiddler, and B.~Zeidler, ``Cyclic code shift
  keying: a low probability of intercept communication technique,'' \emph{IEEE
  Transactions on Aerospace and Electronic Systems}, vol.~39, no.~3, pp.
  786--798, 2003.

\bibitem{mccormick2017simultaneous}
P.~M. McCormick, B.~Ravenscroft, S.~D. Blunt, A.~J. Duly, and J.~G. Metcalf,
  ``Simultaneous radar and communication emissions from a common aperture, part
  ii: experimentation,'' in \emph{2017 IEEE Radar Conference
  (RadarConf)}.\hskip 1em plus 0.5em minus 0.4em\relax IEEE, 2017, pp.
  1697--1702.

\bibitem{liao2011qos}
W.-C. Liao, T.-H. Chang, W.-K. Ma, and C.-Y. Chi, ``{QoS}-based transmit
  beamforming in the presence of eavesdroppers: An optimized
  artificial-noise-aided approach,'' \emph{IEEE Transactions on Signal
  Processing}, vol.~59, no.~3, pp. 1202--1216, 2011.

\bibitem{shafiee2007towards}
S.~Shafiee, N.~Liu, and S.~Ulukus, ``Towards the secrecy capacity of the
  {Gaussian} {MIMO} wire-tap channel: The 2-2-1 channel,'' \emph{arXiv preprint
  arXiv:0709.3541}, 2007.

\bibitem{oggier2007secrecy}
F.~Oggier and B.~Hassibi, ``The secrecy capacity of the {MIMO} wiretap
  channel,'' \emph{arXiv preprint arXiv:0710.1920}, 2007.

\bibitem{ekrem2011secrecy}
E.~Ekrem and S.~Ulukus, ``The secrecy capacity region of the {Gaussian} {MIMO}
  multi-receiver wiretap channel,'' \emph{IEEE Transactions on Information
  Theory}, vol.~57, no.~4, pp. 2083--2114, 2011.

\bibitem{goel2008guaranteeing}
S.~Goel and R.~Negi, ``Guaranteeing secrecy using artificial noise,''
  \emph{IEEE transactions on wireless communications}, vol.~7, no.~6, pp.
  2180--2189, 2008.

\bibitem{negi2005secret}
R.~Negi and S.~Goel, ``Secret communication using artificial noise,'' in
  \emph{IEEE Vehicular Technology Conference}, vol.~62, no.~3.\hskip 1em plus
  0.5em minus 0.4em\relax Citeseer, 2005, p. 1906.

\bibitem{zhang2013design}
X.~Zhang, X.~Zhou, and M.~R. McKay, ``On the design of artificial-noise-aided
  secure multi-antenna transmission in slow fading channels,'' \emph{IEEE
  Transactions on Vehicular Technology}, vol.~62, no.~5, pp. 2170--2181, 2013.

\bibitem{5708173}
J.~P. {Vilela}, M.~{Bloch}, J.~{Barros}, and S.~W. {McLaughlin}, ``Wireless
  secrecy regions with friendly jamming,'' \emph{IEEE Transactions on
  Information Forensics and Security}, vol.~6, no.~2, pp. 256--266, June 2011.

\bibitem{chu2015secrecy}
Z.~Chu, K.~Cumanan, Z.~Ding, M.~Johnston, and S.~Y. Le~Goff, ``Secrecy rate
  optimizations for a {MIMO} secrecy channel with a cooperative jammer,''
  \emph{IEEE Transactions on Vehicular Technology}, vol.~64, no.~5, pp.
  1833--1847, 2015.

\bibitem{vaka2016location}
P.~R. Vaka, S.~Bhattarai, and J.-M. Park, ``Location privacy of non-stationary
  incumbent systems in spectrum sharing,'' in \emph{2016 IEEE Global
  Communications Conference (GLOBECOM)}.\hskip 1em plus 0.5em minus 0.4em\relax
  IEEE, 2016, pp. 1--6.

\bibitem{dimas2017spectrum}
A.~Dimas, B.~Li, M.~Clark, K.~Psounis, and A.~Petropulu, ``Spectrum sharing
  between radar and communication systems: Can the privacy of the radar be
  preserved?'' in \emph{2017 51st Asilomar Conference on Signals, Systems, and
  Computers}.\hskip 1em plus 0.5em minus 0.4em\relax IEEE, 2017, pp.
  1285--1289.

\bibitem{deligiannis2018secrecy}
A.~Deligiannis, A.~Daniyan, S.~Lambotharan, and J.~A. Chambers, ``Secrecy rate
  optimizations for {MIMO} communication radar,'' \emph{IEEE Transactions on
  Aerospace and Electronic Systems}, vol.~54, no.~5, pp. 2481--2492, 2018.

\bibitem{chalise2018performance}
B.~K. Chalise and M.~G. Amin, ``Performance tradeoff in a unified system of
  communications and passive radar: A secrecy capacity approach,''
  \emph{Digital Signal Processing}, vol.~82, pp. 282--293, 2018.

\bibitem{8314727}
K.~{Shen} and W.~{Yu}, ``Fractional programming for communication systems¡ªpart
  i: Power control and beamforming,'' \emph{IEEE Transactions on Signal
  Processing}, vol.~66, no.~10, pp. 2616--2630, May 2018.

\bibitem{wajid2009robust}
I.~Wajid, Y.~C. Eldar, and A.~Gershman, ``Robust downlink beamforming using
  covariance channel state information,'' in \emph{2009 IEEE International
  Conference on Acoustics, Speech and Signal Processing}.\hskip 1em plus 0.5em
  minus 0.4em\relax IEEE, 2009, pp. 2285--2288.

\bibitem{doi:10.1002/ett.4460100604}
\BIBentryALTinterwordspacing
E.~Telatar, ``Capacity of multi-antenna {Gaussian} channels,'' \emph{European
  Transactions on Telecommunications}, vol.~10, no.~6, pp. 585--595. [Online].
  Available:
  \url{https://onlinelibrary.wiley.com/doi/abs/10.1002/ett.4460100604}
\BIBentrySTDinterwordspacing

\bibitem{6626661}
K.~{Cumanan}, Z.~{Ding}, B.~{Sharif}, G.~Y. {Tian}, and K.~K. {Leung},
  ``Secrecy rate optimizations for a {MIMO} secrecy channel with a
  multiple-antenna eavesdropper,'' \emph{IEEE Transactions on Vehicular
  Technology}, vol.~63, no.~4, pp. 1678--1690, May 2014.

\bibitem{4516997}
D.~R. {Fuhrmann} and G.~{San Antonio}, ``Transmit beamforming for {MIMO} radar
  systems using signal cross-correlation,'' \emph{IEEE Transactions on
  Aerospace and Electronic Systems}, vol.~44, no.~1, pp. 171--186, January
  2008.

\bibitem{dinkelbach1967nonlinear}
W.~Dinkelbach, ``On nonlinear fractional programming,'' \emph{Management
  science}, vol.~13, no.~7, pp. 492--498, 1967.

\bibitem{wang2014outage}
K.-Y. Wang, A.~M.-C. So, T.-H. Chang, W.-K. Ma, and C.-Y. Chi, ``Outage
  constrained robust transmit optimization for multiuser {MISO} downlinks:
  Tractable approximations by conic optimization,'' \emph{IEEE Transactions on
  Signal Processing}, vol.~62, no.~21, pp. 5690--5705, 2014.

\bibitem{lobo1998applications}
M.~S. Lobo, L.~Vandenberghe, S.~Boyd, and H.~Lebret, ``Applications of
  second-order cone programming,'' \emph{Linear algebra and its applications},
  vol. 284, no. 1-3, pp. 193--228, 1998.

\bibitem{4350230}
J.~{Li} and P.~{Stoica}, ``{MIMO} radar with colocated antennas,'' \emph{IEEE
  Signal Processing Magazine}, vol.~24, no.~5, pp. 106--114, Sep. 2007.

\bibitem{6884062}
F.~{Wang}, X.~{Wang}, and Y.~{Zhu}, ``Transmit beamforming for multiuser
  downlink with per-antenna power constraints,'' in \emph{2014 IEEE
  International Conference on Communications (ICC)}, June 2014, pp. 4692--4697.

\bibitem{6671453}
A.~{Zappone}, P.~{Cao}, and E.~A. {Jorswieck}, ``Energy efficiency optimization
  in relay-assisted {MIMO} systems with perfect and statistical {CSI},''
  \emph{IEEE Transactions on Signal Processing}, vol.~62, no.~2, pp. 443--457,
  Jan 2014.

\bibitem{law2017optimal}
K.~L. Law, I.~Wajid, and M.~Pesavento, ``Optimal downlink beamforming for
  statistical {CSI} with robustness to estimation errors,'' \emph{Signal
  Processing}, vol. 131, pp. 472--482, 2017.

\bibitem{grant2008cvx}
M.~Grant, S.~Boyd, and Y.~Ye, ``Cvx: Matlab software for disciplined convex
  programming,'' 2008.

\end{thebibliography}

\end{document}